\documentclass[twocolumn,twocolappendix]{aastex631}
\usepackage{amsmath}
\usepackage{graphicx}
\usepackage{upgreek}
\usepackage{enumerate}
\usepackage{natbib}
\usepackage{hyperref}

\newcommand{\oii}{[O\,II] }

\newcommand{\M}{M_{\ast}}

\usepackage{amsmath}

\received{XXX}
\revised{YYY}
\accepted{ZZZ}

\submitjournal{ApJ}

\shorttitle{Galaxy-halo connections for ELGs and LRGs in DESI One-Percent survey}
\shortauthors{Gao et al.}

\begin{document}

\title{The DESI One-Percent survey: constructing galaxy-halo connections for ELGs and LRGs using auto and cross correlations}

\author{Hongyu Gao}
\affil{Department of Astronomy, School of Physics and Astronomy, Shanghai Jiao Tong University, Shanghai, 200240, People's Republic of China}

\author[0000-0002-4534-3125]{Y.P. Jing}
\affil{Department of Astronomy, School of Physics and Astronomy, Shanghai Jiao Tong University, Shanghai, 200240, People's Republic of China}
\affil{Tsung-Dao Lee Institute, and Shanghai Key Laboratory for Particle Physics and Cosmology, Shanghai Jiao Tong University, Shanghai, 200240, People's Republic of China}

\author{Shanquan Gui}
\affil{Department of Astronomy, School of Physics and Astronomy, Shanghai Jiao Tong University, Shanghai, 200240, People's Republic of China}

\author[0000-0002-7697-3306]{Kun Xu}
\affil{Department of Astronomy, School of Physics and Astronomy, Shanghai Jiao Tong University, Shanghai, 200240, People's Republic of China}
\affil{Institute for Computational Cosmology, Durham University, South Road, Durham DH1 3LE, UK}

\author[0000-0001-6575-0142]{Yun Zheng}
\affiliation{Department of Astronomy, School of Physics and Astronomy, Shanghai Jiao Tong University, Shanghai, 200240, People's Republic of China}

\author{Donghai Zhao}
\affil{Department of Astronomy, School of Physics and Astronomy, Shanghai Jiao Tong University, Shanghai, 200240, People's Republic of China}
\affil{Key Laboratory for Research in Galaxies and Cosmology, Shanghai Astronomical Observatory, Shanghai, 200030, People's Republic of China}

\author{Jessica Nicole Aguilar}
\affil{Lawrence Berkeley National Laboratory, 1 Cyclotron Road, Berkeley, CA 94720, USA}

\author[0000-0001-6098-7247]{Steven Ahlen}
\affil{Physics Dept., Boston University, 590 Commonwealth Avenue, Boston, MA 02215, USA}

\author{David Brooks}
\affil{Department of Physics \& Astronomy, University College London, Gower Street, London, WC1E 6BT, UK}

\author{Todd Claybaugh}
\affil{Lawrence Berkeley National Laboratory, 1 Cyclotron Road, Berkeley, CA 94720, USA}

\author{Kyle Dawson}
\affil{Department of Physics and Astronomy, The University of Utah, 115 South 1400 East, Salt Lake City, UT 84112, USA}

\author{Axel de la Macorra}
\affil{Instituto de F\'{\i}sica, Universidad Nacional Aut\'{o}noma de M\'{e}xico, Cd. de M\'{e}xico C.P. 04510, M\'{e}xico}

\author{Peter Doel}
\affil{Department of Physics \& Astronomy, University College London, Gower Street, London, WC1E 6BT, UK}

\author[0000-0003-2371-3356]{Kevin Fanning}
\affil{Center for Cosmology and AstroParticle Physics, The Ohio State University, 191 West Woodruff Avenue, Columbus, OH 43210, USA}
\affil{Department of Physics, The Ohio State University, 191 West Woodruff Avenue, Columbus, OH 43210, USA}

\author{Jaime E. Forero-Romero}
\affil{Departamento de F\'isica, Universidad de los Andes, Cra. 1 No. 18A-10, Edificio Ip, CP 111711, Bogot\'a, Colombia}

\author[0000-0003-3142-233X]{Satya Gontcho A Gontcho}
\affil{Lawrence Berkeley National Laboratory, 1 Cyclotron Road, Berkeley, CA 94720, USA}

\author{Julien Guy}
\affil{Lawrence Berkeley National Laboratory, 1 Cyclotron Road, Berkeley, CA 94720, USA}

\author{Klaus Honscheid}
\affil{Center for Cosmology and AstroParticle Physics, The Ohio State University, 191 West Woodruff Avenue, Columbus, OH 43210, USA}
\affil{Department of Physics, The Ohio State University, 191 West Woodruff Avenue, Columbus, OH 43210, USA}

\author{Robert Kehoe}
\affil{Department of Physics, Southern Methodist University, 3215 Daniel Avenue, Dallas, TX 75275, USA}

\author[0000-0003-1838-8528]{Martin Landriau}
\affil{Lawrence Berkeley National Laboratory, 1 Cyclotron Road, Berkeley, CA 94720, USA}

\author[0000-0003-4962-8934]{Marc Manera}
\affil{Institut de F\'{i}sica d’Altes Energies (IFAE), The Barcelona Institute of Science and Technology, Campus UAB, 08193 Bellaterra Barcelona, Spain}

\author[0000-0002-1125-7384]{Aaron Meisner}
\affil{NSF's NOIRLab, 950 N. Cherry Ave., Tucson, AZ 85719, USA}

\author{Ramon Miquel}
\affil{Institut de F\'{i}sica d’Altes Energies (IFAE), The Barcelona Institute of Science and Technology, Campus UAB, 08193 Bellaterra Barcelona, Spain}
\affil{Instituci\'{o} Catalana de Recerca i Estudis Avan\c{c}ats, Passeig de Llu\'{\i}s Companys, 23, 08010 Barcelona, Spain}

\author[0000-0002-2733-4559]{John Moustakas}
\affil{Department of Physics and Astronomy, Siena College, 515 Loudon Road, Loudonville, NY 12211, USA}

\author[0000-0001-8684-2222]{Jeffrey A. Newman}
\affil{Department of Physics \& Astronomy and Pittsburgh Particle Physics, Astrophysics, and Cosmology Center (PITT PACC), University of Pittsburgh, 3941 O'Hara Street, Pittsburgh, PA 15260, USA}

\author[0000-0001-6590-8122]{Jundan Nie}
\affil{National Astronomical Observatories, Chinese Academy of Sciences, A20 Datun Rd., Chaoyang District, Beijing, 100012, P.R. China}

\author[0000-0002-0644-5727]{Will Percival}
\affil{Department of Physics and Astronomy, University of Waterloo, 200 University Ave W, Waterloo, ON N2L 3G1, Canada}
\affil{Perimeter Institute for Theoretical Physics, 31 Caroline St. North, Waterloo, ON N2L 2Y5, Canada}
\affil{Waterloo Centre for Astrophysics, University of Waterloo, 200 University Ave W, Waterloo, ON N2L 3G1, Canada}

\author{Graziano Rossi}
\affil{Department of Physics and Astronomy, Sejong University, Seoul, 143-747, Korea}

\author{Michael Schubnell}
\affil{Department of Physics, University of Michigan, Ann Arbor, MI 48109, USA}

\author[0000-0002-6588-3508]{Hee-Jong Seo}
\affil{Department of Physics \& Astronomy, Ohio University, Athens, OH 45701, USA}

\author[0000-0003-1704-0781]{Gregory Tarl\'{e}}
\affil{University of Michigan, Ann Arbor, MI 48109, USA}

\author{Benjamin Alan Weaver}
\affil{NSF's NOIRLab, 950 N. Cherry Ave., Tucson, AZ 85719, USA}

\author{Jiaxi Yu}
\affil{Ecole Polytechnique F\'{e}d\'{e}rale de Lausanne, CH-1015 Lausanne, Switzerland}

\author[0000-0002-4135-0977]{Zhimin Zhou}
\affil{National Astronomical Observatories, Chinese Academy of Sciences, A20 Datun Rd., Chaoyang District, Beijing, 100012, P.R. China}

%\author{more authors...}

\correspondingauthor{Y.P. Jing}
\email{ypjing@sjtu.edu.cn}

\begin{abstract}
In the current Dark Energy Spectroscopic Instrument (DESI) survey, emission line galaxies (ELGs) and luminous red galaxies (LRGs) are essential for mapping the dark matter distribution at $z \sim 1$. We measure the auto and cross correlation functions of ELGs and LRGs at $0.8<z\leq 1.0$ from the DESI One-Percent survey. Following Gao et al. (2022), we construct the galaxy-halo connections for ELGs and LRGs simultaneously. With the stellar-halo mass relation (SHMR) for the whole galaxy population (i.e. normal galaxies), LRGs can be selected directly by stellar mass, while ELGs can also be selected randomly based on the observed number density of each stellar mass, once the probability $P_{\mathrm{sat}}$ of a satellite galaxy becoming an ELG is determined. We demonstrate that the observed small scale clustering prefers a halo mass-dependent $P_{\mathrm{sat}}$ model rather than a constant. With this model, we can well reproduce the auto correlations of LRGs and the cross correlations between LRGs and ELGs at $r_{\mathrm{p}}>0.1$ $\mathrm{Mpc}\,h^{-1}$. We can also reproduce the auto correlations of ELGs at $r_{\mathrm{p}}>0.3$ $\mathrm{Mpc}\,h^{-1}$ ($s>1$ $\mathrm{Mpc}\,h^{-1}$) in real (redshift) space. Although our model has only seven parameters, we show that it can be extended to higher redshifts and reproduces the observed auto correlations of ELGs in the whole range of $0.8<z<1.6$, which enables us to generate a lightcone ELG mock for DESI. With the above model, we further derive halo occupation distributions (HODs) for ELGs which can be used to produce ELG mocks in coarse simulations without resolving subhalos. 
\end{abstract}


\section{Introduction} \label{sec:intro}
A precise understanding of the connection between galaxies and dark matter is one of the most critical challenges in current research. Galaxies are formed in dark matter halos, and the growth of galaxies is closely related to the growth of their host halos \citep{2018ARA&A..56..435W}. Thus, establishing the galaxy-halo connection is a prerequisite for understanding galaxy formation and evolution. Moreover, cosmological probes, such as baryon acoustic oscillation \citep[BAO, e.g.,][]{2005MNRAS.362..505C,2005ApJ...633..560E}, redshift-space distortion \citep[RSD, e.g.,][]{1987MNRAS.227....1K}, and weak gravitational lensing \citep[e.g.,][]{2001PhR...340..291B, 2018ARA&A..56..393M} provide us with powerful ways to infer the cosmological parameters and constrain the dark energy model. To make these cosmological probes accurate enough to fulfill the requirements of current cosmological studies, an accurate relation between the galaxies and the underlying dark matter halos is needed.

In recent years, a series of statistical approaches have been proposed to construct the galaxy-halo connection. For instance, in classical halo occupation distribution (HOD) framework \citep[e.g.,][]{1998ApJ...494....1J, 2000MNRAS.318.1144P, 2000ApJ...543..503M, 2000MNRAS.318..203S, 2002ApJ...575..587B, 2005ApJ...633..791Z, 2007ApJ...667..760Z, 2011ApJ...736...59Z, 2015MNRAS.454.1161Z, 2016MNRAS.457.4360Z, 2018MNRAS.476.1637Z, 2016MNRAS.459.3040G, 2016MNRAS.460.1173R, 2016MNRAS.460.3647X, 2018MNRAS.481.5470X, 2018MNRAS.478.2019Y, 2019ApJ...879...71W, 2020MNRAS.497..581A, 2022MNRAS.510.3301Y}, halo occupation number $\left \langle N\left(M\right) \right \rangle$ is used to quantify the mean number of galaxies hosted by a halo with mass $M$. Additionally, conditional luminosity function (CLF) or conditional stellar mass function (CSMF) \citep{2003MNRAS.339.1057Y, 2006MNRAS.365..842C, 2007MNRAS.376..841V, 2009ApJ...695..900Y, 2012ApJ...752...41Y, 2015ApJ...799..130R, 2018ApJ...858...30G, 2020MNRAS.499..631V} can present more detailed physical properties of galaxies in a given halo. Furthermore, by making full use of the physical information of subhalos, abundance matching (SHAM) technique can link galaxies in observations to halos and subhalos in simulations \citep[e.g.,][]{1998ApJ...506...19W, 2006MNRAS.371..537W, 2006MNRAS.371.1173V, 2010ApJ...717..379B, 2010MNRAS.402.1796W, 2010MNRAS.404.1111G, 2012MNRAS.423.3458S, 2013MNRAS.428.3121M, 2014MNRAS.437.3228G, 2016MNRAS.459.3040G, 2016MNRAS.460.3100C, 2018ARA&A..56..435W, 2019MNRAS.488.3143B, 2022ApJ...926..130X, 2022MNRAS.516...57Y, 2022ApJ...939..104X, 2022ApJ...925...31X, 2023ApJ...944..200X}. This method has become one of the most efficient ways to determine the stellar-halo mass relation (SHMR) and the galaxy stellar mass function (SMF).

Galaxy samples complete to a stellar mass (or a broad-band luminosity) are typically required in previous studies of the galaxy-halo connection. These samples are usually constructed, with proper incompleteness corrections, from flux-limited redshift surveys such as the Sloan Digital Sky Survey \citep[SDSS,][]{2000AJ....120.1579Y, 2006AJ....131.2332G}, the VIMOS VLT
Deep Survey \citep[VVDS,][]{2000AJ....120.1579Y, 2006AJ....131.2332G}, the Deep
Extragalactic Evolutionary Probe 2 \citep[DEEP2,][]{2013ApJS..208....5N} and the VIMOS Public Extragalactic Redshift Survey \citep[VIPERS,][]{2014A&A...566A.108G,2014A&A...562A..23G,2018A&A...609A..84S}. Since these galaxies represent the general population, we refer to them as {\textit{normal galaxies}} in this paper. Above a certain stellar mass threshold, these normal galaxies are complete, and thus their clustering and SMF can be used to constrain the SHMR. However, at medium and high redshifts, due to the limited wavelength coverage and detection depth, galaxies are usually color and magnitude selected, such as the SDSS-III
Baryon Oscillation Spectroscopic Survey \citep[BOSS,][]{2013AJ....145...10D} and Dark Energy Spectroscopic Instrument \citep[DESI,][]{2013arXiv1308.0847L, 2016arXiv161100036D,2016arXiv161100037D,2022AJ....164..207D} surveys, which makes it challenging to construct the stellar mass (or luminosity) limited samples. Recently, Photometric objects
Around Cosmic webs \citep[PAC,][]{2022ApJ...925...31X} method has been proposed to overcome this difficulty. Utilizing the correlations between the photometric data from the DESI Legacy Imaging
Survey \citep{2019AJ....157..168D} and spectroscopic samples at various redshifts, \cite{2023ApJ...944..200X} greatly improved the SHMR measurements down to $10^{8.0} M_{\odot}$ at $ z\sim 0.2$ and $10^{9.8} M_{\odot}$ at $z \sim 0.7$. Nevertheless, at $z\sim 1$, PAC still requires a large sample of galaxies with deeper photometric observations over the wide sky area of a redshift survey.

As galaxy surveys are extended to higher redshifts, galaxies with specific spectral features have become the main targets of current spectroscopic surveys, such as SDSS IV
extended Baryon Oscillation Spectroscopic Survey \citep[eBOSS,][]{2016AJ....151...44D}, DESI \citep{2013arXiv1308.0847L, 2016arXiv161100036D,2016arXiv161100037D,2022AJ....164..207D} and Subaru Prime Focus Spectrograph \citep[PFS,][]{2014PASJ...66R...1T,2022SPIE12184E..10T}. In particular, DESI achieves coverage of the sky area over 14,000 $\mathrm{deg^2}$, and is devoted to targeting more than 8 million luminous red galaxies (LRGs) at $0.4<z<1.0$ and 16 million \oii emission line galaxies (ELGs) at $0.6<z<1.6$. The combination of these two types of galaxy samples will provide us with an invaluable opportunity to study the galaxy-halo connection at $z \sim 1$. However, it is a big challenge to accurately model the galaxy-halo connection for these targeted galaxies and constrain the overall SHMR, since they are incomplete for stellar mass limits. The incompleteness could be very complicated due to their complex color and magnitude selection especially for ELGs (see \cite{2023AJ....165..126R}). Different from the normal galaxy population, many studies \citep[e.g.,][]{2012MNRAS.426..679G, 2013MNRAS.432.2717C, 2016MNRAS.461.3421F, 2017MNRAS.472..550F,2018MNRAS.474.4024G, 2019ApJ...871..147G, 2020MNRAS.498.1852G, 2020MNRAS.499.5486A, 2021MNRAS.502.3599H, 2021MNRAS.505.2784Z, 2021PASJ...73.1186O, 2022MNRAS.512.5793Y, 2022arXiv221010068H, 2022arXiv221010072H, 2023MNRAS.519.4253L} have proved that the HOD of ELGs is expected to peak at some host halo mass ($\sim10^{12}\,M_{\odot}$) and decrease as the host halo mass increases, but the HOD forms obtained vary widely.  

More and more studies focus on improving the galaxy-halo connection models for LRG and ELG samples at $z > 0.5$. For example, by jointly modeling the BOSS LRGs and eBOSS ELGs, \cite{2019ApJ...871..147G} adopted the incomplete conditional stellar mass function (ICSMF) model to constrain the completeness of ELG, the galaxy quenched fraction and the SHMR down to $\sim 10^{10}\,M_{\odot}$. For efficient HOD analysis, \cite{2022MNRAS.512.5793Y} developed a multi-tracer HOD framework that can model the cross correlations of LRGs, ELGs and Quasars (QSOs) as well as their environment-based secondary galaxy bias. But the combination of different HOD models for the three galaxy tracers introduces a large number of parameters, which increases the difficulty of the precise constraints of the parameters. Using the DESI-like mock samples from a hydrodynamical simulation, \cite{2022arXiv221010068H, 2022arXiv221010072H} optimized the HODs for LRG and ELG through modeling galaxy conformity effects and improved the theoretical predictions of clustering on both small and large scales, though the current generation of hydrodynamical simulations still lack the power to accurately predict the properties of galaxies as required by the current precise cosmological studies. As we will show later, for the complicated target selection, the HOD of ELGs in DESI has a so complicated dependence on redshift that it is extremely challenging to propose an analytical expression, in contrast to that for normal galaxies.

Instead of using the conventional HOD approach, \cite{2022ApJ...928...10G} (hereafter Paper I) has proposed a novel SHAM approach to construct the SHMR and the galaxy-halo connection for ELGs. They measured the auto and cross correlations between ELGs and the stellar mass-selected normal galaxy samples from VIPERS. They determined the SHMR for normal galaxies using the abundance and clustering of stellar mass-selected samples. They then proposed that ELGs could be randomly selected from the normal galaxy population, as long as the ELG satellite fraction is reasonably reduced, and the satellite fraction changes with the strength of the \oii emission line. They demonstrate that this approach can well reproduce the auto and cross correlations for ELGs in both real-space and redshift-space. The main advantage of this approach is that the fundamental relation between the galaxy and the host halo (or subhalo) is determined by stellar mass, and only the satellite fraction is a free parameter for ELGs. Other studies \citep[e.g.,][]{2015ApJ...799..130R,2021ApJ...919...25W,2022A&A...663A..85Z} also implied that SHMR for ELGs is similar to that for normal galaxies, since the star-forming galaxies dominate the whole population at the relevant stellar mass range ($\M <10^{10.5} M_\odot$) and the host halo mass only weakly depends on galaxy color when stellar mass is fixed (see in particular Figure 10 of \cite{2015ApJ...799..130R}).

The LRG and ELG samples with high spectroscopic completeness from the One-Percent survey of the DESI Survey Validation \citep[SV,][]{sv} enable us to further refine the SHAM method in Paper I and to extend the galaxy-halo connection model to redshift $z \sim 1$. On the one hand, by combining the LRG and ELG samples, we can probe the SHMR at both the low and high mass ends. LRGs dominate the high mass end ($\M > 10^{11} \, M_{\odot}$) of the SMF, and ELGs are star-forming blue galaxies with low and intermediate mass ($\M < 10^{10.5} \, M_{\odot}$). These two types of galaxies cover a wide range of stellar masses. On the other hand, although ELGs have a wide range of host halo masses, most central ELGs are expected to be located in small halos with $M_{\mathrm{h}}<10^{12.5}M_\odot$. The LRGxELG cross-correlation can help to reveal the distribution of ELGs around the massive halos, that is, the distribution of ELGs as satellites in massive halos. Therefore, with the cross-correlations of the overlapping LRG and ELG samples, in addition to their auto correlations, we can achieve a stronger constraint on the ELG-halo connection. 

In this work, we will first measure the auto and cross correlations of the LRG and ELG samples from the One-Percent survey, as well as their observed number densities. Then, following Paper I, we will simultaneously determine both the SHMR for normal galaxies and the ELG-halo connection. We will demonstrate that after modeling the normal galaxies in simulation using the SHMR, LRGs can be selected from the massive normal galaxies, while ELGs can be selected randomly from the normal galaxies based on the observed number density after reducing the fraction of the satellite galaxies, which is a function of the host halo mass. With our models, we will develop a method to generate an ELG mock sample that has the same number density and clustering properties as the DESI ELG sample. We expect that the mock samples will be very useful for future cosmological studies based on the DESI ELGs. Finally, it is straightforward to generate mock samples for the DESI LRG sample, since the target selection criterion is relatively simple.

The paper is structured as follows. In Section \ref{sec:data}, we describe the observed galaxy samples and show our measurements of galaxy clustering. In Section \ref{sec:modeling}, we present our basic ideas for modeling the galaxy-halo connection using the N-body simulation. The fitting results are presented in Section \ref{sec:results}. In Section \ref{sec:hod_of_ELG}, we derive the HOD for ELG based on our ELG-halo connection. Finally, we briefly summarize the main results of this work. The cosmological parameters used in the calculations and simulations in this paper are $\Omega_{\mathrm{m},0} = 0.268$, $\Omega_{\Lambda,0} = 0.732$ and $H_0 = 100h \,\mathrm{km\, s^{-1}\,Mpc^{-1}}=71 \,\mathrm{km\,s^{-1}\,Mpc^{-1}}$.
 
Our work is one of many studies for the galaxy-halo connections in the One-Percent survey. Other relevant parallel studies include: HOD modeling for ELGs \citep{abacusELG_Rocher,ELG_Lasker}, for LRGs \citep{LRG_Ereza,abacusLRGQSO_Yuan}, for QSOs \citep{QSO_Rajeev,abacusLRGQSO_Yuan}, BGS HOD \citep{BGS_Grove,BGS_Smith}, and for multi-tracers\citep{multi-tracerHOD_Yuan}, and SHAM modeling for the different tracers \citep{inclusiveSHAM,overviewSHAM}. The combined efforts will greatly improve the current understanding of the galaxy-halo connection for the different tracers in DESI.

\begin{figure}
	\centering
	\includegraphics[scale=0.6]{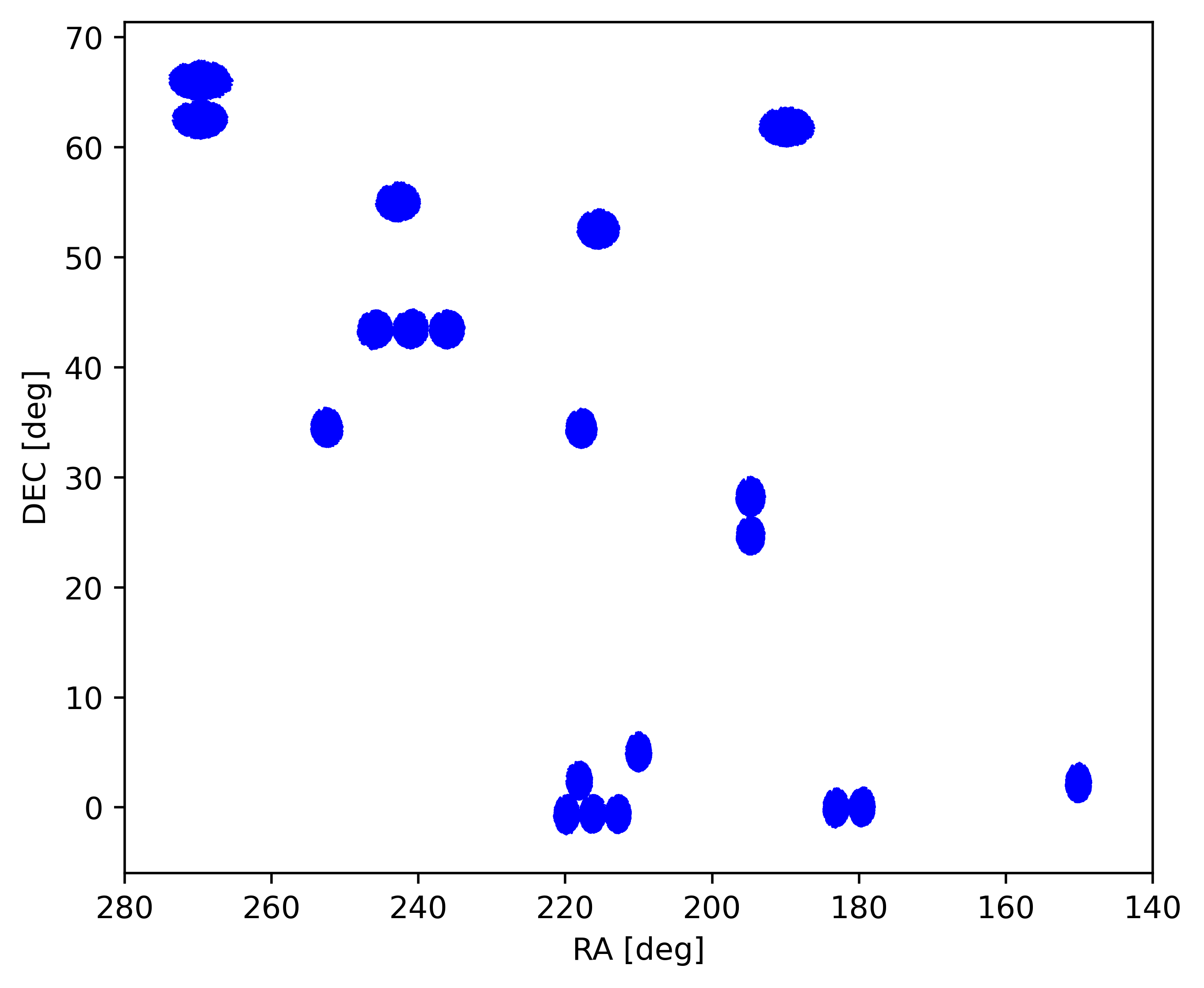}
	\caption{Sky coverage of the DESI One-Percent survey.
		\label{fig:sky coverage}}
\end{figure}

\begin{figure*}
	\centering
	\includegraphics[scale=0.8]{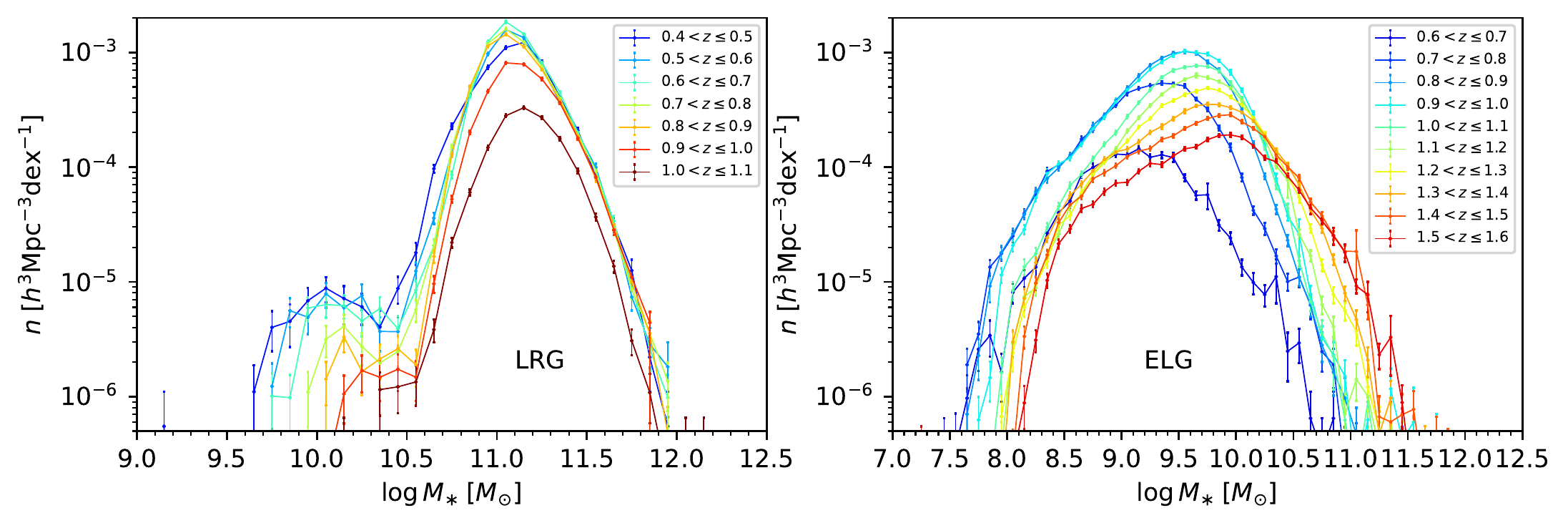}
	\caption{Evolution of the SMFs of the LRG and ELG samples in the One-Percent survey. The left and right panels correspond to LRG and ELG respectively. The data points with Poisson errors denote the observed SMFs in different redshift bins. In the measurements, each galaxy has been multiplied by the completeness weight as mentioned in Section \ref{subsec:sv3}.
		\label{fig:lrg_and_elg_number_density}}
\end{figure*}

\begin{figure}
	\centering
	\includegraphics[scale=0.8]{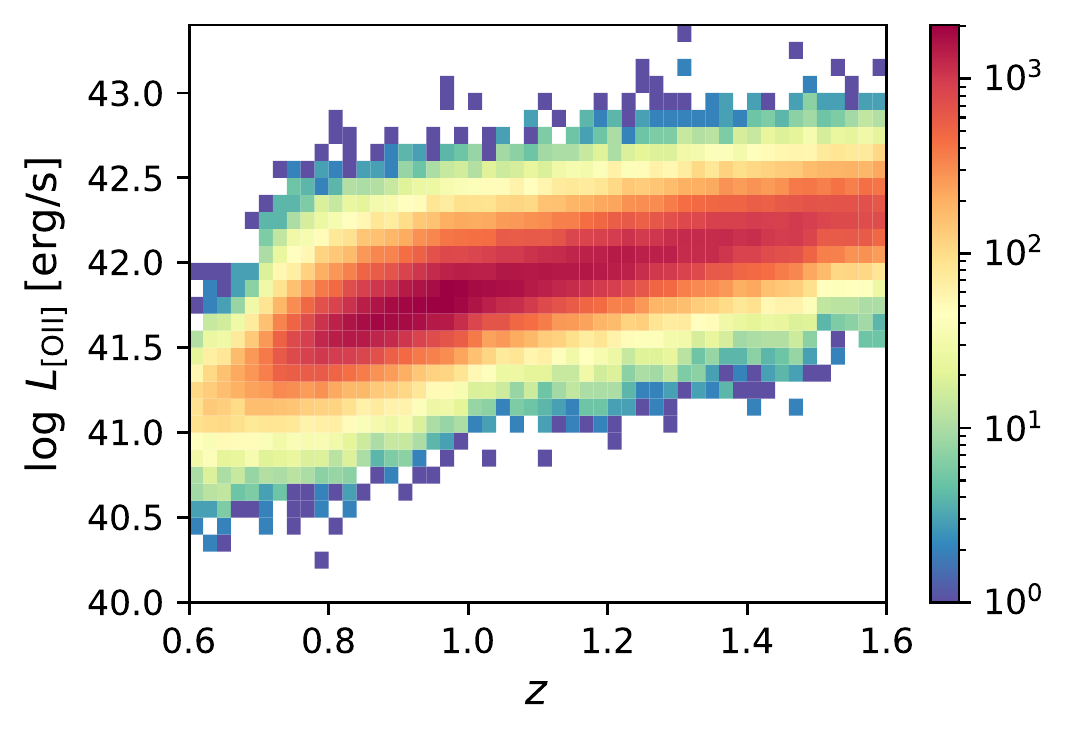}
	\caption{Two-dimensional distribution of \oii luminosity and redshift for ELG samples in the One-Percent survey. The color bar represents the count of galaxies in each tiny grid.
		\label{fig:oii}}
\end{figure}

\begin{deluxetable}{cccc}
	\tablenum{1}
	\tablecaption{Details of four LRG subsamples. }
	\label{tab:lrg}
	\tablehead{ \colhead{Name} &
		\colhead{Redshift Range} & \colhead{$\log \M\, [M_{\odot}]$ } &\colhead{$N_{\mathrm{g}}$}
	}
	\startdata
	LRG0&$0.8<z\leq 1.0$& $\left[11.1, 11.3\right]$& 13906 \\
	LRG1&$0.8<z\leq 1.0$& $\left[11.3, 11.5\right]$& 4834 \\
	LRG2&$0.8<z\leq 1.0$& $\left[11.5, 11.7\right]$& 957 \\
	LRG3&$0.8<z\leq 1.0$& $\left[11.7, 11.9\right]$& 124 \\
	\enddata
	%\tablecomments{}
\end{deluxetable}

\begin{deluxetable}{cccc}
	\tablenum{2}
	\tablecaption{Details of four ELG subsamples.}
	\label{tab:elg}
	\tablehead{ \colhead{Name} &
		\colhead{Redshift Range} & \colhead{$\log \M\, [M_{\odot}]$ } &\colhead{$N_{\mathrm{g}}$}
	}
	\startdata
	ELG0&$0.8<z\leq 1.0$& $\left[8.5, 9.0\right]$& 9481 \\
	ELG1&$0.8<z\leq 1.0$& $\left[9.0, 9.5\right]$& 29764 \\
	ELG2&$0.8<z\leq 1.0$& $\left[9.5, 10.0\right]$& 34155 \\
	ELG3&$0.8<z\leq 1.0$& $\left[10.0, 10.5\right]$& 6583 \\
	\enddata
	%\tablecomments{}
\end{deluxetable}

\begin{figure*}
	\centering
	\includegraphics[scale=0.6]{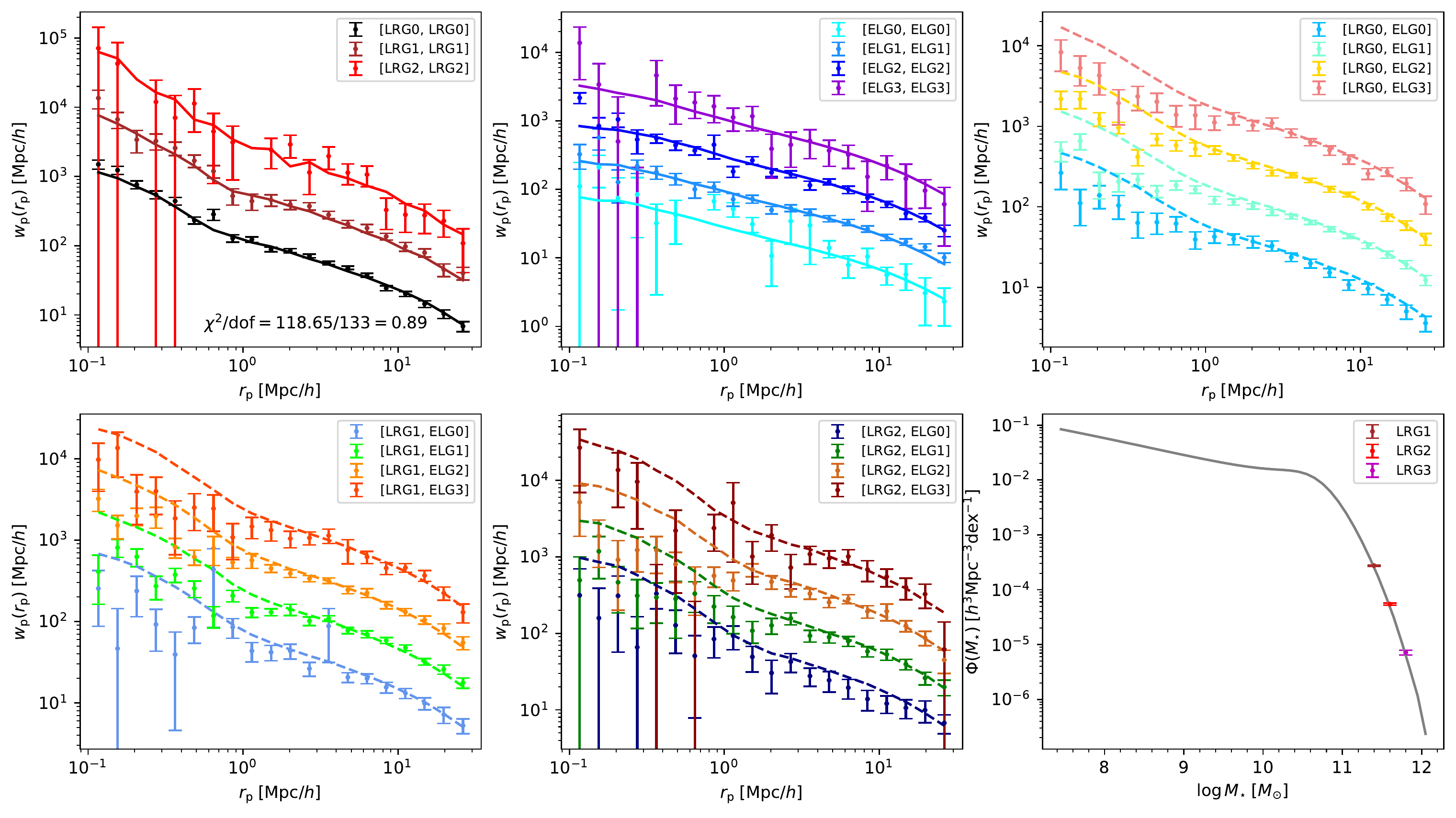}
	\caption{Projected cross (auto) correlation functions of LRG and ELG subsamples at $0.8<z\leq1.0$, and fitting results for the constant $P_{\mathrm{sat}}$ model with $\boldsymbol{w}^{\mathrm{obs}}_{\mathrm{p},\mathrm{LRG}}$, $\boldsymbol{w}^{\mathrm{obs}}_{\mathrm{p},\mathrm{ELG}}$ and $n^{\mathrm{obs}}_{\mathrm{LRG}}$. Data points with error bars in the first five panels represent the observed $\boldsymbol{w}^{\mathrm{obs}}_{\mathrm{p}}$, and different colors indicate different combinations of subsamples. The solid lines with the corresponding colors indicate the best-fit models for $\boldsymbol{w}^{\mathrm{mod}}_{\mathrm{p}}$ while the dashed lines denote the model predictions (not fitting). The last panel exhibits the observed number densities of LRG subsamples as well as the modeled SMF $\Phi(\M)$ measured in simulation. The reduced $\chi^2$ marked on the top left panel is calculated using only the statistics being fitted (see also Equation \ref{equ:chi2_all_auto}). To make a clear presentation, each $\boldsymbol{w}_{\mathrm{p}}$ has been multiplied by a factor of $3^n$, where $n$ is taken as 0, 1, 2 and 3 from the bottom one to the top one (except for the top left panel in which $n$ changes from 0 to 2).
		\label{fig:wp_auto}}
\end{figure*}

\begin{figure*}
	\centering
	\includegraphics[scale=0.6]{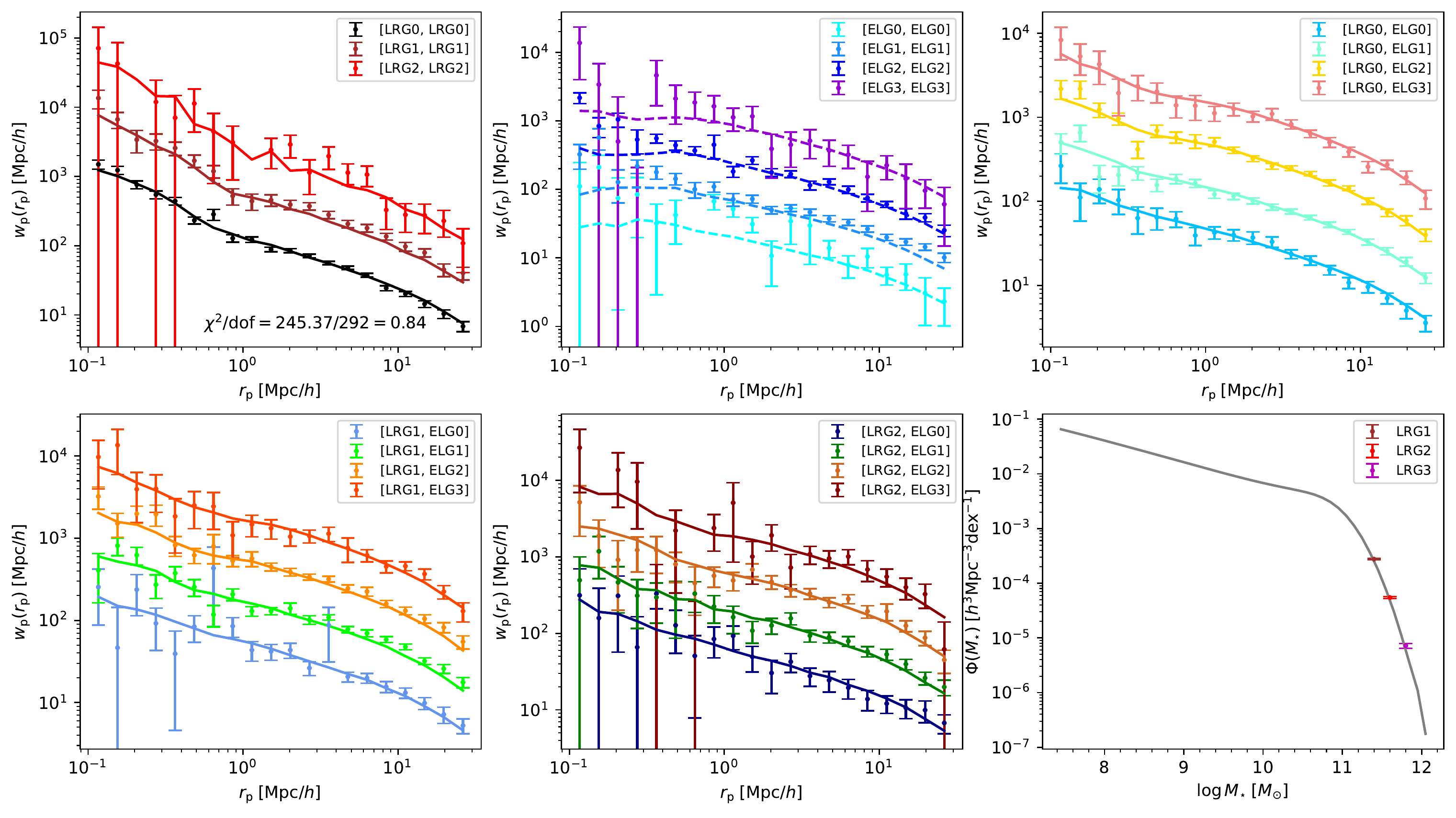}
	\caption{Similar to Figure \ref{fig:wp_auto}, but fitting results for the constant $P_{\mathrm{sat}}$ model with $\boldsymbol{w}^{\mathrm{obs}}_{\mathrm{p},\mathrm{LRG}}$, $\boldsymbol{w}^{\mathrm{obs}}_{\mathrm{p},\mathrm{LRGxELG}}$ and $n^{\mathrm{obs}}_{\mathrm{LRG}}$. The reduced $\chi^2$ marked on the top left panel is calculated using only the statistics being fitted (see also Equation \ref{equ:chi2_all_cross}).
		\label{fig:wp_cross}}
\end{figure*}

\begin{figure*}
	\centering
	\includegraphics[scale=0.6]{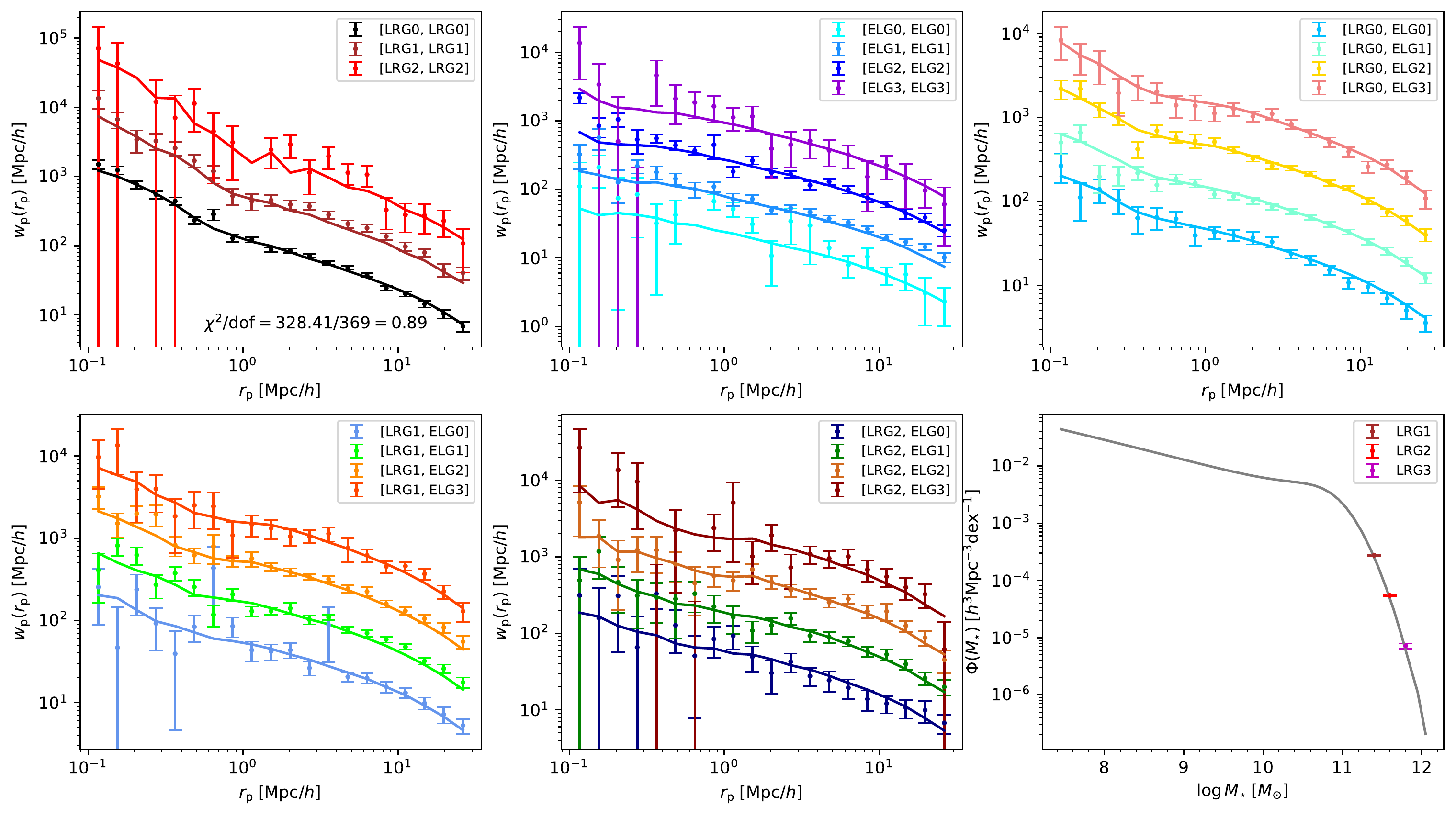}
	\caption{Similar to Figure \ref{fig:wp_auto} and Figure \ref{fig:wp_cross}, but fitting results for the halo mass-dependent $P_{\mathrm{sat}}$ model with all the cross (auto) correlations $\boldsymbol{w}^{\mathrm{obs}}_{\mathrm{p},\mathrm{LRG}}$, $\boldsymbol{w}^{\mathrm{obs}}_{\mathrm{p},\mathrm{ELG}}$ and $\boldsymbol{w}^{\mathrm{obs}}_{\mathrm{p},\mathrm{LRGxELG}}$, and the LRG number densities $n^{\mathrm{obs}}_{\mathrm{LRG}}$. The reduced $\chi^2$ marked on the top left panel is calculated using all the statistics being fitted (see also Equation \ref{equ:chi2_all}).
		\label{fig:wp}}
\end{figure*}

\section{Galaxy samples and clustering measurements} \label{sec:data}
In this section, we briefly introduce the One-Percent survey. We describe the selection of LRG and ELG subsamples, and show their observed number densities as a function of stellar mass and redshift. The measurements of the auto and cross correlation functions are also presented.

\subsection{DESI One-Percent survey} \label{subsec:sv3}
DESI is dedicated to collecting the spectra for approximately 40 million extra-galactic objects covering more than 14,000 $\mathrm{deg^2}$ in five years \citep{2013arXiv1308.0847L, 2016arXiv161100036D,2016arXiv161100037D,2022AJ....164..207D}. The spectroscopic observations of DESI are performed by a multi-object, fiber-fed spectrograph attached to the prime focus panel of the 4-meter Mayall telescope at Kitt Peak National Observatory \citep{2022AJ....164..207D}. The spectrograph spans a wavelength range of $3600-9800\, \AA$ and can assign fibers to 5,000 objects at a time \citep{2016arXiv161100037D, 2023AJ....165....9S, corrector}. The multiple supporting pipelines of DESI experiment are described in detail by \cite{2023AJ....165..144G,redrock2023,fba,ops,2023AJ....165...50M}. The target selections and survey validations of DESI can be found in a series of papers \citep{2020RNAAS...4..188A, 2020RNAAS...4..187R, 2020RNAAS...4..181Z, 2020RNAAS...4..180R, 2020RNAAS...4..179Y, 2023ApJ...943...68L, 2023AJ....165..124A, 2023ApJ...947...37C, 2023AJ....165..253H, 2023ApJ...944..107C, 2023AJ....165..126R, 2023AJ....165...58Z}. The parent catalog used for DESI target selections is constructed from Data Release 9 of the DESI Legacy
Imaging Surveys \citep{2017PASP..129f4101Z,2019AJ....157..168D, dr9}. The photometric data contains three optical bands $grz$ from the DECam Legacy Survey \citep[DECaLS,][]{2019AJ....157..168D}, the Dark Energy Survey \citep[DES,][]{2005astro.ph.10346T}, the Beijing-Arizona Sky Survey \citep[BASS,][]{2017PASP..129f4101Z} and the Mayall z-band Legacy
Survey (MzLS), and two infrared bands $W1W2$ from the Wide-field Infrared Survey Explorer \citep[WISE,][]{2010AJ....140.1868W}. Four bands $grzW1$ are used to select the DESI LRG targets in $0.4<z<1.0$ \citep{2023AJ....165...58Z}, while a $g$-band magnitude cut and a $grz$ color cut are designed to select the DESI ELG targets in $0.6<z<1.6$ \citep{2023AJ....165..126R}.

The One-Percent survey (also known as the 1 \% survey) is the final stage of DESI Survey Validation \citep[SV,][]{sv}. It was operated from 5 April 2021 to 10 June 2021, covering 20 separate "rosette" areas, each of which is approximately 7 $ \mathrm{\deg^{2}}$. Thus, the total sky area of the One-Percent survey is about 1 \% of the main survey. The One-Percent survey adopts the same observing mode as the main survey, but it performs spectroscopic measurements for all potential targets as much as possible by conducting many repeated visits. Because of the high fiber-assignment and spectroscopic rate, the galaxy samples in the One-Percent survey are nearly complete. The Sky coverage of the One-Percent survey is displayed in Figure \ref{fig:sky coverage}.
 
%We use Early Data Assembly version 1 (EDAv1) of the SV3 LSS 
We use the One-Percent survey LSS clustering catalog, which is part of the DESI Early Data Release \citep[EDR,][]{dr}. The LSS catalog contains all target classes with successful redshift measurements from the internal "Fuji" spectroscopic data releases. The completeness weight for each galaxy is estimated by performing 128 alternative Merged Target List (MTL) realizations. The assignment probability PROB of a target can be calculated by $N_{\mathrm{assigned}}/N_{\mathrm{tot}}$, where $N_{\mathrm{tot}} = 129$ is the total number of realizations and $N_{\mathrm{assigned}}$ indicates the number of times a target is assigned in these 129 realizations. Then the completeness weight $w_{\mathrm{c}}$ is defined as $129/(128\times \mathrm{PROB}+1)$. 
By comparing the ELG auto correlation functions with and without angular upweighting \citep{2020MNRAS.498..128M}, we find that this weighting can boost the $\boldsymbol{w}_{\mathrm{p}}$ by about 5\% at $r_{\mathrm{p}} \sim 0.5$ $\mathrm{Mpc}\,h^{-1}$ and by about 10\% at $r_{\mathrm{p}} \sim 0.1$ $\mathrm{Mpc}\,h^{-1}$. Given the relatively large uncertainty of the ELG auto correlations on small scales, the effect of the angular upweighting on our results is negligible. 
%We have checked that the effect of angular upweighting \citep{2020MNRAS.498..128M} on our results is not significant, due to the high completeness of fiber-assignment and relatively large measurement errors of correlation functions on small scales in the One-Percent survey.

\subsection{LRG and ELG samples} \label{subsec:samples}
Taking the photometry data in five bands $grzW1W2$, we perform spectral energy distribution (SED) fitting for the LRG and ELG samples using {\tt\string CIGALE} \citep{2019A&A...622A.103B}. We adopt the stellar spectral library provided by \cite{2003MNRAS.344.1000B} to construct stellar population synthesis models. The initial stellar mass function (IMF) provided by \cite{2003PASP..115..763C} is used in the calculation. We suppose three different metallicities $Z/Z_{\odot} = 0.4, 1.0, 2.5$ in our model. A delayed star formation history (SFH) $\phi(t) \simeq t \exp(-t/\tau)$ is assumed, where the
timescale $\tau$ spans from $10^7$ to $1.258 \times 10^{10}$ yr with an equal logarithmic space of 0.1 dex. We apply the starburst reddening law of \cite{2000ApJ...533..682C} to calculate the dust attenuation, in which the color excess $E(B-V)$ varies from 0 to 0.5.

We calculate the number density of galaxies as functions of the stellar mass and redshift, which are presented in Figure \ref{fig:lrg_and_elg_number_density}. They are the stellar mass functions (SMFs) of the observed samples. One should be aware that the observed SMFs may suffer from various target selections, and they should be distinguished from the intrinsic ones. Nevertheless, the figure tells us the basic properties of the samples. First, for the LRG sample, its SMF hardly changes between redshifts 0.4 and 1.0 at the massive end $ M_{\ast} > 10^{11.3}M_{\odot}$. This is also consistent with previous studies \citep{2007A&A...474..443P, 2010A&A...523A..13P, 2013A&A...558A..23D, 2022ApJ...939..104X,2023ApJ...944..200X}. At the massive end, \cite{2023ApJ...944..200X} demonstrated that the SMF has nearly no evolution since $z<0.7$. Here we consider the massive LRG with $ M_{\ast} > 10^{11.3}M_{\odot}$ as a stellar mass-complete sample. At the low mass end ($M_{\ast} < 10^{10.3}M_{\odot}$), the SMF of LRG shows a second peak. We find that these galaxies near this peak are dusty star-forming galaxies with large $E(B-V)$ (about 0.4-0.5). We have also looked at the individual 5-band magnitudes, and found that most of these galaxies are too bright in $W2$ relative to the LRG SED expectation, which also prefers dusty star-forming SEDs. Moreover, this population is rather small ($<1$ \%) and mostly at low redshifts ($0.4<z<0.6$), therefore our results in this paper are not affected by these galaxies.

As for the ELG, the change in the SMF is more remarkable from $z=0.6$ to $z=1.6$. The number density of ELG reaches a maximum at $z \sim 1$ and then decreases with increasing redshift. At fixed redshift, the SMF of ELG shows a peak between $10^{9} M_{\odot}$ and $10^{10} M_{\odot}$. On the one hand, the ELG targets in DESI are mainly blue galaxies with intense star formation and strong emission lines. Massive galaxies are usually more prone to quenching and reddening, resulting in a decrease in the ELG number density at the massive end. On the other hand, some low-mass faint galaxies could be excluded in target selection, leading to a drop in the SMF at the low-mass end. Nevertheless, due to the low mass-to-light ratio of these blue galaxies, there is still a considerable number of ELGs even at $10^{8.5} M_{\odot}$. The changes of the ELG SMF with redshift also reflect the complexity of its target selection. Combining the massive LRGs and the low-mass ELGs, our galaxy samples can cover a wide range of stellar mass.

In addition to stellar mass, we also present the distribution of redshift and \oii luminosity for the ELG samples in Figure \ref{fig:oii}. The \oii fluxes are taken from the EDAv1 catalog. We notice that some bright ELGs are missing at $z<0.8$, which may be attributed to the bright cut of $g>20$ mag in the ELG target selection \citep{2023AJ....165..126R}. This cut intends to reduce the contamination of galaxies at low redshifts. Therefore, we first focus on the redshift range $0.8 < z \leq 1.0$, where the ELG has a more complete population and achieves the highest number density, and the LRG sample is also complete.

To investigate the cross correlation between LRG and ELG, we divide our LRG and ELG samples at $0.8<z\leq 1.0$ into different subsamples (LRG0, LRG1, LRG2, LRG3, ELG0, ELG1, ELG2, ELG3) binned by stellar mass. The number of galaxies and the stellar mass range of each subsample are presented in Table \ref{tab:lrg} and \ref{tab:elg}. It should be emphasized that LRG0 may be somewhat incomplete in stellar mass, so we only use its auto and cross correlation functions instead of its number density in the subsequent fitting process. Besides, since the number of galaxies in LRG3 is limited, its clustering signal is severely affected by Poisson noise. We do not include the correlation functions of LRG3 in our fitting.

\subsection{Estimation of galaxy correlation function} \label{subsec:clustering}
For subsamples $x$ and $y$, their two-dimensional correlation function can be estimated via the Landy-Szalay estimator \citep{1993ApJ...412...64L, 1998ApJ...494L..41S}
\begin{equation}
\xi_{xy}\left(r_{\mathrm{p}}, r_{\mathrm{\pi}}\right) = \left[ \frac{D_xD_y-D_xR_y-D_yR_x+R_xR_y}{R_xR_y}\right],
\end{equation}
where $r_{\mathrm{p}}$ and $r_{\pi}$ correspond to the two components that are perpendicular and parallel to the line-of-sight, and $D_xD_y$, $D_xR_y$, $D_yR_x$ and $R_xR_y$ are the normalized weighted pair counts for galaxy-galaxy, galaxy-random, random-galaxy and random-random, respectively. A total of 20 equally logarithmic $r_{\mathrm{p}}$ bins from $0.1$ to $30$ $\mathrm{Mpc}\,h^{-1}$ and 40 equally linear $r_{\pi}$ bins from $0$ to $40$ $\mathrm{Mpc}\,h^{-1}$ are set in the measurement. We use all the random samples provided by the One-Percent survey catalog to account for the survey geometry. We keep the RA and Dec of these random samples but assign them new redshifts by shuffling the observed redshifts of each subsample.

To obtain the real-space projected correlation function $w_{\mathrm{p},xy} \left(r_{\mathrm{p}}\right)$, we integrate $\xi_{xy} \left(r_{\mathrm{p}}, r_{\mathrm{\pi}}\right)$ along the line-of-sight direction \citep{1983ApJ...267..465D} by
\begin{equation}
w_{\mathrm{p},xy}\left(r_{\rm p}\right) = 2\int_{0}^{r_{\pi, \rm max}} \xi_{xy} \left( r_{\rm p}, r_{\pi} \right)\mathrm{d}r_{\pi},
\end{equation}
where $r_{\pi, \rm max}=40 \,\mathrm{Mpc}\,h^{-1}$. 

Utilizing the jackknife technique, we can quantify the covariance matrix of the observed correlation functions. We divide the survey areas of One-Percent survey into 100 approximately equal small fields according to the distribution of random points. The covariance matrix is computed by
\begin{equation}
C_{ij} = \frac{N_{\mathrm{jack}}-1}{N_{\mathrm{jack}}} \sum_{k=1}^{N_{\mathrm{jack}}} \left( w^k_{\mathrm{p}, i} - \bar{w}_{\mathrm{p}, i} \right) \left( w^k_{\mathrm{p}, j} - \bar{w}_{\mathrm{p}, j} \right), \label{equ:covariance}
\end{equation}
where $N_{\mathrm{jack}}=100$ and $w^k_{\mathrm{p}, i(j)}$ is measured from the $k$-th jackknife region, here $i$ ($j$) represents the $i$ ($j$)-th $r_\mathrm{p}$ bin. 

The observed $\boldsymbol{w}_{\mathrm{p}}$ are presented as data points with error bars in Figure \ref{fig:wp_auto}. 

Similar to the measurement of $\xi_{xy}\left(r_{\mathrm{p}}, r_{\mathrm{\pi}}\right)$, we also calculate the correlation functions $\xi_{xy}\left(s, \mu \right)$ in redshift-space and express them as multipole moments \citep{1992ApJ...385L...5H}
\begin{equation}
\xi_{l,xy}\left(s\right)=\frac{2l+1}{2}\int_{-1}^{1}\xi_{xy}\left(s,\mu\right)L_l\left(\mu\right) \mathrm{d}\mu,
\end{equation}
where $L_l\left(\mu\right)$ is the Legendre function, and $l$ can be specified as 0 (monopole), 2 (quadrupole) and 4 (hexadecapole). We take 15 equally logarithmic $s$ bins from $0.3$ to $30$ $\mathrm{Mpc}\,h^{-1}$ and 10 equally linear $\mu$ bins from 0 to 1 in the measurements.

\section{Modeling} \label{sec:modeling}
In this section, we propose a concise abundance matching technique to connect the observed LRG and ELG samples to dark matter halos in N-body simulation. 

\subsection{N-body simulation} \label{subsec:simulation}

We adopt the {\tt\string CosmicGrowth} \citep{2019SCPMA..6219511J} simulation suite in this study. The {\tt\string CosmicGrowth} is performed by an adaptive parallel P$^3$M algorithm \citep{2002ApJ...574..538J}, and releases a series of N-body simulations with different cosmological parameters and different resolutions. We choose one of the $\Lambda$CDM simulations of {\tt\string CosmicGrowth} to model the galaxy-halo connection. This simulation adopts the standard cosmology: $\Omega_{\mathrm{m}} = 0.268$, $\Omega_{\Lambda} = 0.732$, $h=0.71$, $n_{\mathrm{s}}=0.968$ and $\sigma_8=0.83$, and has a total of $3072^3$ dark matter particles in a $600\, \mathrm{Mpc}\,h^{-1}$ box, which yields a high mass resolution of $m_{\mathrm{p}} = 5.54 \times 10^8 \,M_{\odot}\,h^{-1}$. The halo groups are identified by the friends-of-friends algorithm (FOF) \citep{1985ApJ...292..371D} with a linking length of $b=0.2$, while the subhalo merger trees are built using the Hierarchical-Bound-Tracing algorithm ({\tt\string HBT+}) \citep{2012MNRAS.427.2437H,2018MNRAS.474..604H}. 

For the subhalos with less than 20 particles, we calculate their merger time scales using the fitting formula provided by \cite{2008ApJ...675.1095J} and exclude those subhalos that have fully merged with their central subhalos. The halo mass function from \cite{2019SCPMA..6219511J} and the subhalo mass function from \cite{2022ApJ...925...31X} have verified that the halos in this simulation can be well resolved to $10^{10} \, M_{\odot}\,h^{-1}$ (about 20 particles), which is sufficient for modeling ELGs that are more likely to live in low-mass halos \citep[e.g.,][]{2016MNRAS.461.3421F, 2019ApJ...871..147G, 2021MNRAS.502.3599H, 2021PASJ...73.1186O}. The mass of a host halo $M_{\mathrm{h}}$ is defined as its current $M_{\mathrm{vir}}$ that is the mass enclosed by a virialized spherical structure with an over-density $\Delta_{\mathrm{vir}}\left(z\right)$ \citep{1972ApJ...176....1G,1998ApJ...495...80B}. The accretion mass of a subhalo $M_{\mathrm{s}}$ is defined as the $M_{\mathrm{vir}}$ at the snapshot before it was accreted by the current host halo. Moreover, for the calculation of clustering, we define the $z$-axis as the direction of the line-of-sight and add the RSD effects to the coordinate components along this direction for all the halos and subhalos in simulation. To cover the whole redshift range of ELGs in the observation ($0.6<z\leq 1.6$), we take a total of five halo catalogs from snapshots at $z= 0.71$, $0.92$, $1.09$, $1.27$ and $1.47$.

\subsection{Stellar-halo mass relation} \label{subsec:shmr}
To link the dark matter halos in the simulation to the observed galaxies, we use stellar mass as a bridge. For a given halo (subhalo) with mass $M_{\mathrm{h}}$, we assume that the stellar mass of the hosted galaxy follows a Gaussian conditional probability distribution function (PDF) 
\begin{equation}
p(\M|M_{\mathrm{h}}) = \frac{1}{\sqrt{2\pi}\sigma}\exp\left[-\frac{\left(\log \M - \log \left\langle \M|M_{\mathrm{h}} \right\rangle\right)^2}{2\sigma^2}\right], \label{equ:p(M_star|M_h)}
\end{equation}
where $\left\langle \M|M_{\mathrm{h}} \right\rangle$ denotes the mean SHMR and $\sigma$ is the scatter of this relation.
We adopt a double power-law function \citep{2006MNRAS.371..537W,2010MNRAS.402.1796W,2012ApJ...752...41Y,2013MNRAS.428.3121M} to parameterize the $\left\langle \M|M_{\mathrm{h}} \right\rangle$  
\begin{equation}
\left\langle \M|M_{\mathrm{h}} \right\rangle = \frac{2k}{\left(M_{\mathrm{h}}/M_0\right)^{-\alpha} + \left(M_{\mathrm{h}}/M_0\right)^{-\beta}},
\end{equation}
where $M_0$ divides the SHMR into two parts with different slopes $\alpha$ and $\beta$, and $k$ is a normalization constant. Similarly, the relation between stellar mass and accretion mass $p(\M|M_{\mathrm{s}})$ for a subhalo can also be established using the above model. Since the stellar mass of a galaxy can still be influenced by the subsequent evolution process after infalling \citep{2012ApJ...752...41Y}, the SHMR of a subhalo should be expected to be different from that of a halo. However, this difference is small in the modeling of galaxy clustering \citep{2010MNRAS.402.1796W}, especially when the observed sample size is limited. Therefore, similar to other works \citep[e.g.,][]{2010MNRAS.402.1796W,2019MNRAS.488.3143B,2022ApJ...925...31X,2023ApJ...944..200X}, we adopt a unified model with five parameters: $\{\alpha, \beta, M_0, k, \sigma\}$ to describe the $p(\M|M_{\mathrm{h}})$ for halos and $p(\M|M_{\mathrm{s}})$ for subhalos. After the release of DESI Y1 data, we will attempt to construct the SHMR for halos and subhalos separately and test this assumption.

\subsection{Modeling LRGs and ELGs in simulation} \label{subsec:elg-halo}
Once the SHMR is established, we can assign stellar mass to each halo and subhalo. In this way, we can obtain a population of galaxies with complete stellar masses in the simulation. We refer to this kind of galaxy as a normal galaxy. 

Firstly, from the normal galaxies, we can directly select the LRG subsamples (LRG0, LRG1, LRG2, LGR3) according to the ranges of their stellar mass bins. As mentioned in Section \ref{subsec:samples}, compared to LRG1, LRG2 and LRG3, the LRG0 in the observation may be somewhat incomplete. Although we still choose all normal galaxies with $M_{\ast}$ between $10^{11.1}$ and $10^{11.3} M_{\odot}$ as LRG0, we only consider their clustering rather than their number density. 

Next, we improve the approach provided by Paper I to select ELGs. Paper I introduced a constant parameter $f_{\mathrm{sat}}$ to reduce the fraction of satellite galaxies in the simulation, and then randomly sampled ELGs from the simulated normal galaxies based on the ELG number density in observation. Here, we further divide this process into two steps. 

The first step is to select ELG candidates from the normal galaxies. We keep all central galaxies in the normal population as ELG candidates, that is, the probability of the central galaxies to be selected is $P_{\mathrm{cen}}=1$. Since satellite galaxies are more likely to be quenched by the environment, the probability that a satellite galaxy becomes an ELG candidate, denoted by $P_{\mathrm{sat}}$, is usually less than 1. $P_{\mathrm{sat}}$ is determined by best-fitting the auto and cross correlations of ELGs and LRGs, as studied in the next section.  Here in the selection of ELG candidates, we keep all central galaxies and randomly remove some satellite galaxies based on their $P_{\mathrm{sat}}$.

The second step is to select true ELGs from these ELG candidates. For a given stellar mass bin, we can measure the ELG number density $\bar{n}^{\mathrm{obs}}\left(M_{\ast}\right)$ in observation and the candidate number density $\bar{n}^{\mathrm{can}}\left(M_{\ast}\right)$ in simulation. Then, the probability that a candidate is selected as an ELG is written as
\begin{equation}
F^{\mathrm{ELG}}\left(M_{\ast}\right) = \frac{\bar{n}^{\mathrm{obs}}\left(M_{\ast}\right)}{\bar{n}^{\mathrm{can}}\left(M_{\ast}\right)}. \label{equ:f(M_star)}
\end{equation}   
We calculate $F^{\mathrm{ELG}}\left(M_{\ast}\right)$ from $10^{7.5}$ to $10^{12} M_{\odot}$ with a width $\Delta \log \M = 0.1$, and linearly interpolate the $F^{\mathrm{ELG}}-\log M_{\ast}$ relation to make it continuous. In this way, we can assign a probability $F^{\mathrm{ELG}}$ to each candidate and randomly select ELGs based on their $F^{\mathrm{ELG}}$. Finally, the selected ELGs in simulation can be further categorized into four subsamples: ELG0, ELG1, ELG2 and ELG3.

We summarize the modeling process for LRGs and ELGs in simulation as follows:
\begin{enumerate}[(i)]
 \item Populate halos and subhalos with normal galaxies according to the SHMR model $p(\M|M_{\mathrm{h}})$ and $p(\M|M_{\mathrm{s}})$.

 \item Select the four LRG subsamples LRG0, LRG1, LRG2 and LRG3 from the normal galaxies population, and compute their number density $n^{\mathrm{mod}}$.

 \item Given the probability $P_{\mathrm{sat}}$ for satellite galaxies, keep all central galaxies and randomly select some satellite galaxies as ELG candidates based on their $P_{\mathrm{sat}}$.

 \item Calculate the probability $F^{\mathrm{ELG}}$ for each ELG candidate and randomly select some candidates as ELGs based on their $F^{\mathrm{ELG}}$.

 \item Select the four ELG subsamples ELG0, ELG1, ELG2 and ELG3 from the final ELG samples.

 \item Calculate the auto and cross correlation functions for all the LRG and ELG subsamples in simulation.
\end{enumerate}

For each set of parameters of the SHMR and $P_{\mathrm{sat}}$ model, we follow the above steps to calculate the number densities, and  auto and cross correlations of LRGs and ELGs. In order to make the model predictions stable, we implement the above steps 10 times and average the results from the 10 realizations as the final $\boldsymbol{w}^{\mathrm{mod}}_{\mathrm{p}}$ and $n^{\mathrm{mod}}$. The model parameters are determined by best-fitting the model predictions to the observations, as described in the next section.

\begin{deluxetable}{cccc}
	\tablenum{3}
	\tablecaption{Best-fit parameters of the SHMRs for different $P_{\mathrm{sat}}$ models.}
	\label{tab:best-fit}
	\tablehead{ \colhead{} & \colhead{constant $P_{\mathrm{sat}}$} & \colhead{constant $P_{\mathrm{sat}}$}& \colhead{$P_{\mathrm{sat}}(M_{\mathrm{h}})$}
	}
	\startdata
	 $\boldsymbol{w}^{\mathrm{obs}}_{\mathrm{p},\mathrm{LRG}}$&\checkmark&\checkmark&\checkmark\\ 
	 $\boldsymbol{w}^{\mathrm{obs}}_{\mathrm{p},\mathrm{ELG}}$&\checkmark& &\checkmark \\ 
	 $\boldsymbol{w}^{\mathrm{obs}}_{\mathrm{p},\mathrm{LRGxELG}}$& &\checkmark&\checkmark \\ 
	 $n^{\mathrm{obs}}_{\mathrm{LRG}}$&\checkmark&\checkmark&\checkmark \\ 
	 \hline
	 $\log M_0 \, [M_{\odot}\,h^{-1}]$& $11.56^{+0.25}_{-0.22}$&$12.14^{+0.09}_{-0.10}$&$12.07^{+0.09}_{-0.09}$\\ 
	 $\alpha$&$0.43^{+0.01}_{-0.03}$&$0.37^{+0.03}_{-0.04}$&$0.37^{+0.03}_{-0.03}$\\ 
	 $\beta$&$2.72^{+1.33}_{-0.84}$&$2.27^{+0.38}_{-0.24}$&$2.61^{+0.45}_{-0.28}$\\ 
	 $\log k$&$10.11^{+0.14}_{-0.11}$&$10.40^{+0.07}_{-0.06}$&$10.36^{+0.06}_{-0.05}$\\ 
	 $\sigma$&$0.18^{+0.02}_{-0.02}$&$0.21^{+0.01}_{-0.01}$&$0.21^{+0.01}_{-0.01}$\\ 
	 $P_{\mathrm{sat}}$&$0.89^{+0.07}_{-0.11}$&$0.15^{+0.02}_{-0.01}$\\ 
	 $a$& & &$1 \, (\mathrm{fixed})$\\ 
	 $b$& & &$12.55^{+0.09}_{-0.15}$\\ 
	 $c$& & &$0.04^{+0.03}_{-0.03}$\\ 
	\enddata
	\tablecomments{The first two columns represent the constant $P_{\mathrm{sat}}$ model, and the third column denotes the halo mass-dependent $P_{\mathrm{sat}}(M_{\mathrm{h}})$ model. The check marks in the first four rows indicate which observational quantities are used in the fit. The best-fit model parameters as well as 1$\sigma$ uncertainties are shown in the remaining rows.}
\end{deluxetable}

\begin{figure*}
	\centering
	\includegraphics[scale=0.8]{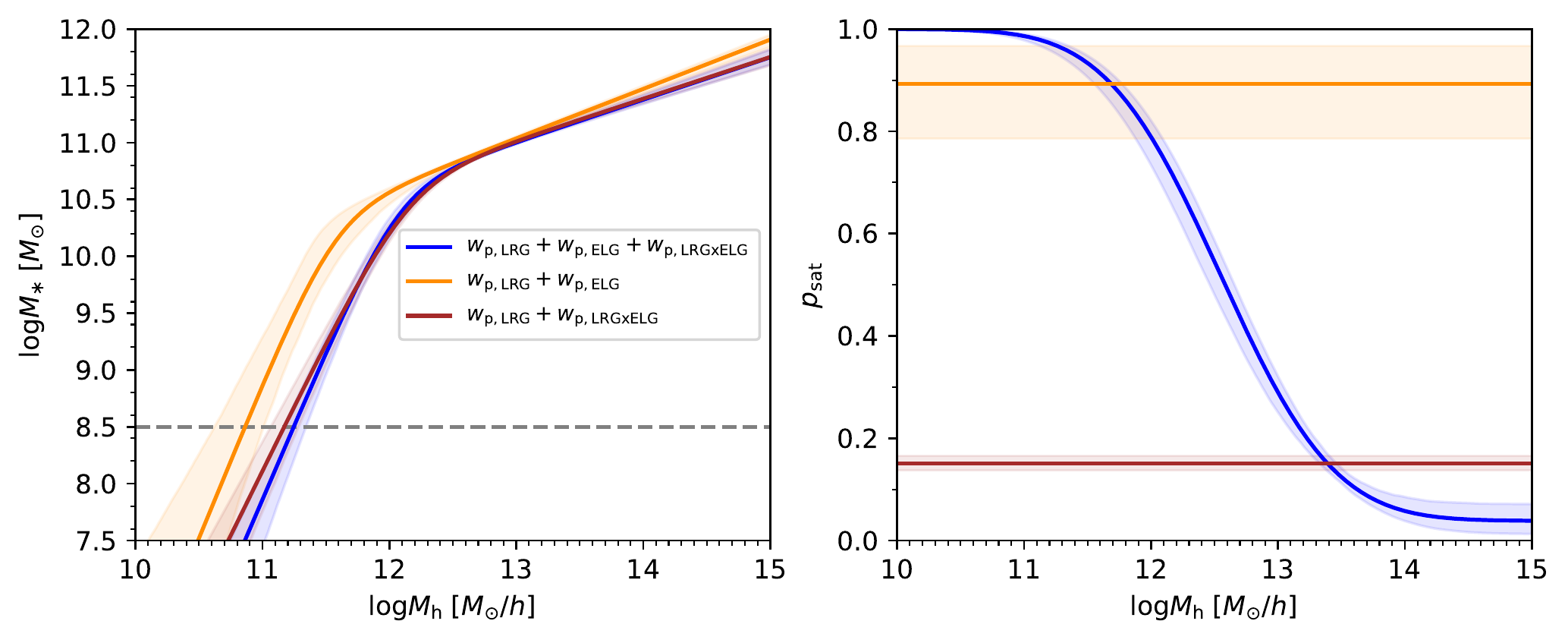}
	\caption{Constraints of SHMR (left panel) and $P_{\mathrm{sat}}$ (right panel) models. The best-fit results (the first two columns of Table \ref{tab:best-fit}) for the constant $P_{\mathrm{sat}}$ model using either ELG auto or LRGxELG cross correlation are displayed as orange and brown curves, respectively. The blue solid lines represent the best-fit result (the last column of Table \ref{tab:best-fit}) for the halo mass-dependent $P_{\mathrm{sat}}$ model using all the correlation functions. The shallow regions denote the $1\sigma$ scatter. The horizontal gray line indicates the lowest stellar mass limit that can be probed by the ELG subsample.
		\label{fig:shmr}}
\end{figure*}

\begin{figure*}
	\centering
	\includegraphics[scale=0.6]{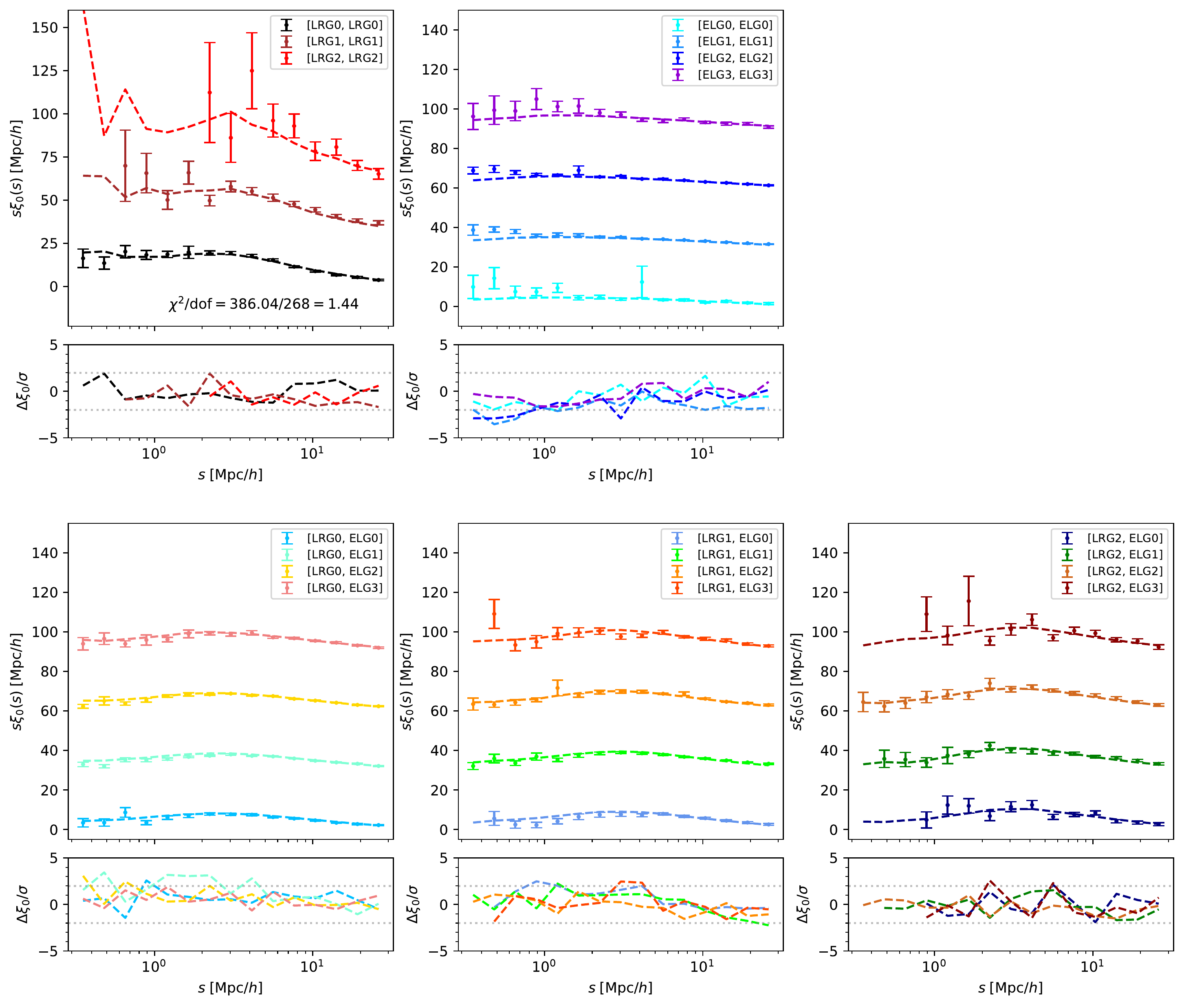}
	\caption{Observed monopole $\xi_0(s)$ in redshift-space and model predictions. The data points with error bars represent the measurements. The different auto and cross correlations of LRGs and ELGs are shown in different panels. The dashed lines are the direct model predictions using the best-fit parameters. To make a clear presentation, each $s\xi_0$ has been added with a constant $n\times 30$, where $n$ is taken as 0, 1, 2 and 3 from the bottom to the top one (except for the top left panel in which $n$ changes from 0 to 2). The deviation $\Delta \xi_0$ of the model from the data divided by the measurement error $\sigma$ is also shown at the bottom of each panel.
		\label{fig:xi0}}
\end{figure*}

\begin{figure*}
	\centering
	\includegraphics[scale=0.6]{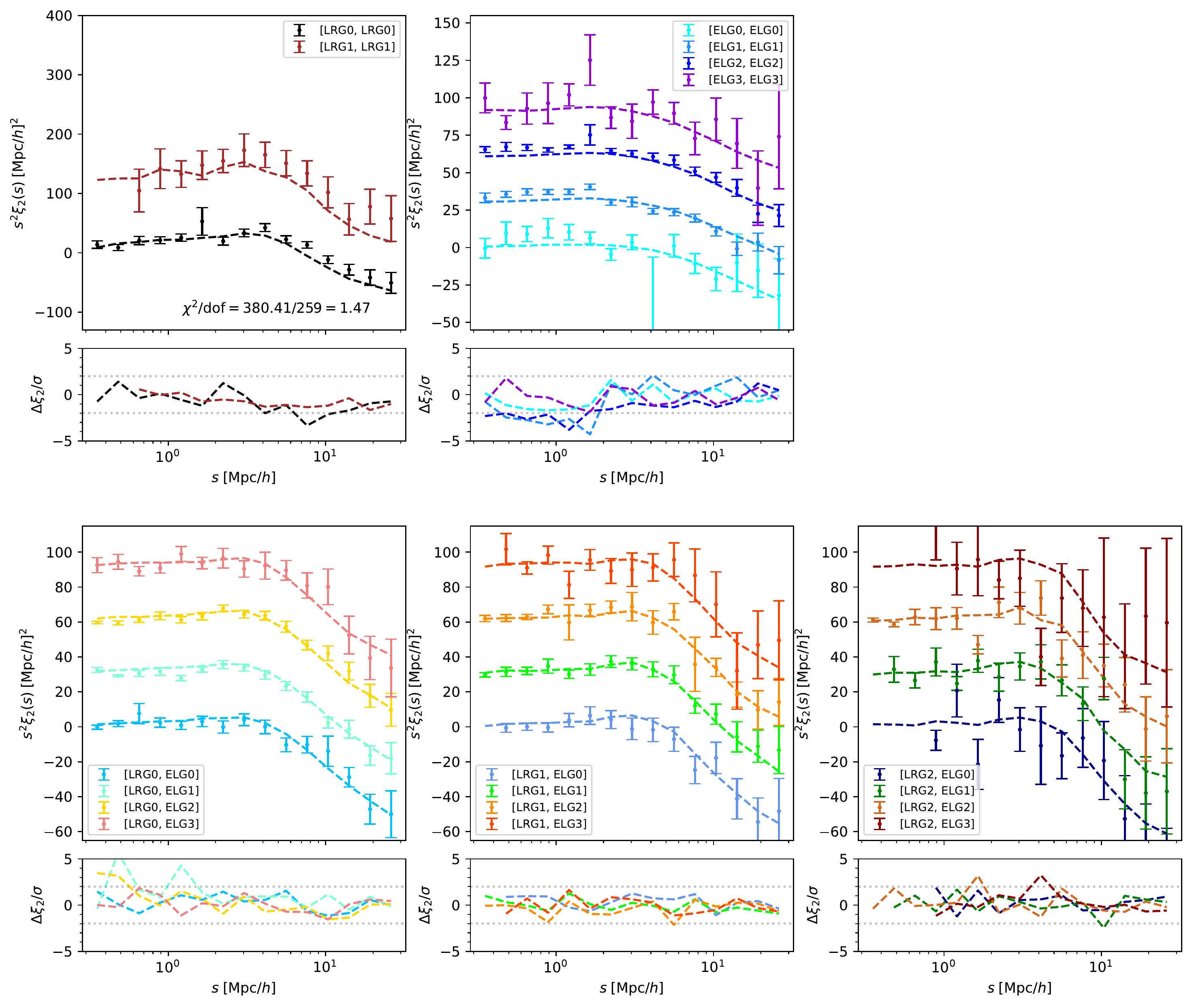}
	\caption{Similar to Figure \ref{fig:xi0}, but we show the observed quadrupole $\xi_2(s)$ in redshift-space and model predictions. To make a clear presentation, each $s^2\xi_2(s)$ has been added with a constant $n\times 30$, where $n$ is taken as 0, 1, 2 and 3 from the bottom one to the top one (except for the top left panel in which each $s^2\xi_2(s)$ has been added with $n\times 100$ where $n$ changes from 0 to 1). Here we omit the $\xi_2(s)$ of the LRG2 auto correlations, because its measurement is too noisy. The deviation $\Delta \xi_2$ of the model from the data divided by the measurement error $\sigma$ is also shown at the bottom of each panel.
		\label{fig:xi2}}
\end{figure*}

\begin{figure*}
	\centering
	\includegraphics[scale=0.6]{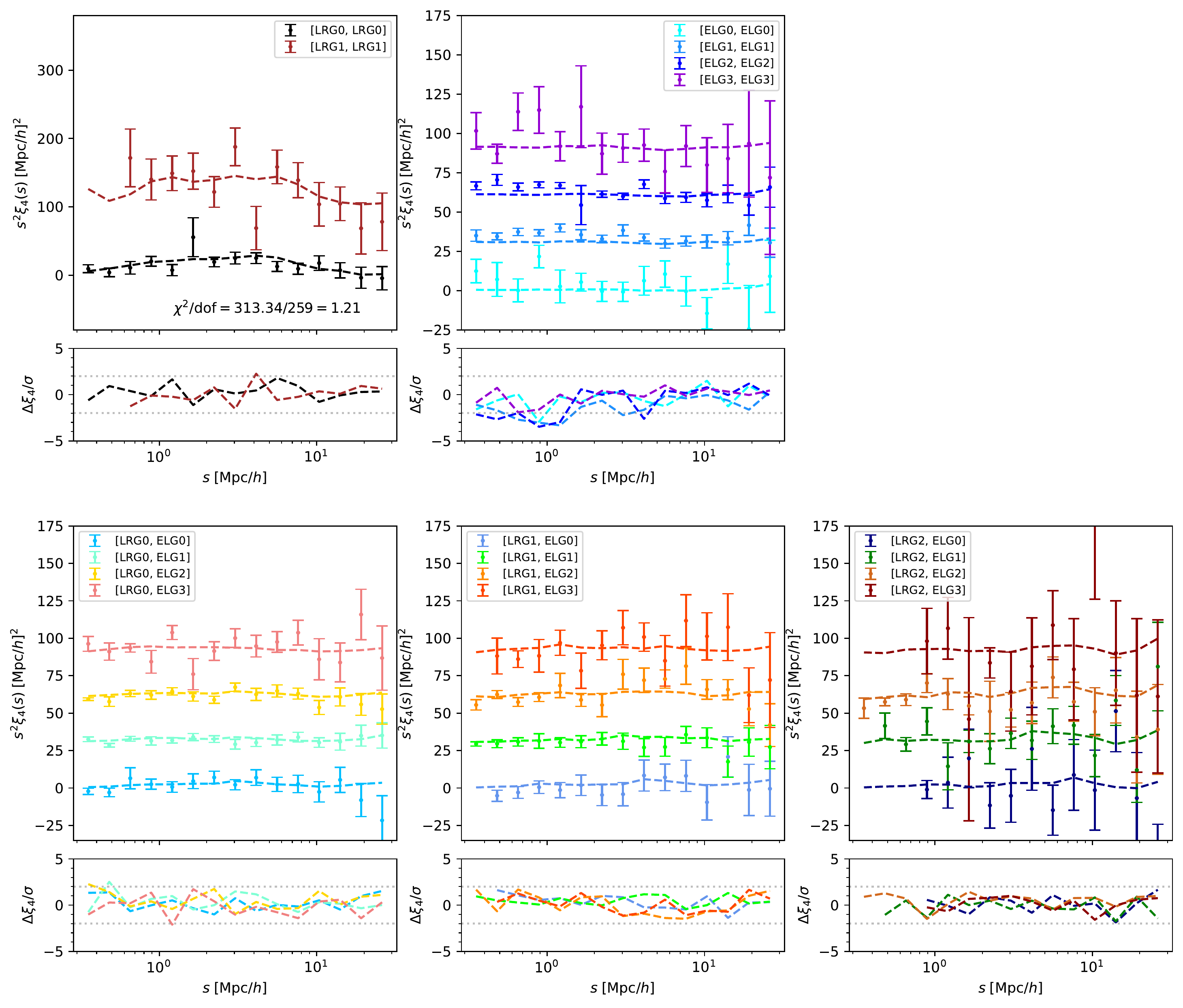}
	\caption{Similar to Figure \ref{fig:xi0}, but we show the observed hexadecapole $\xi_4(s)$ in redshift-space and the model predictions. To make a clear presentation, each $s^2\xi_4(s)$ has been added with a constant $n\times 30$, where $n$ is taken as 0, 1, 2 and 3 from the bottom one to the top one (except for the top left panel in which each $s^2\xi_4(s)$ has been added with $n\times 100$ where $n$ changes from 0 to 1). Here we omit the $\xi_4(s)$ of the LRG2 auto correlations, because its measurement is too noisy. The deviation $\Delta \xi_4$ of the model from the data divided by the measurement error $\sigma$ is also shown at the bottom of each panel.
		\label{fig:xi4}}
\end{figure*}

\begin{figure*}
	\centering
	\includegraphics[scale=0.6]{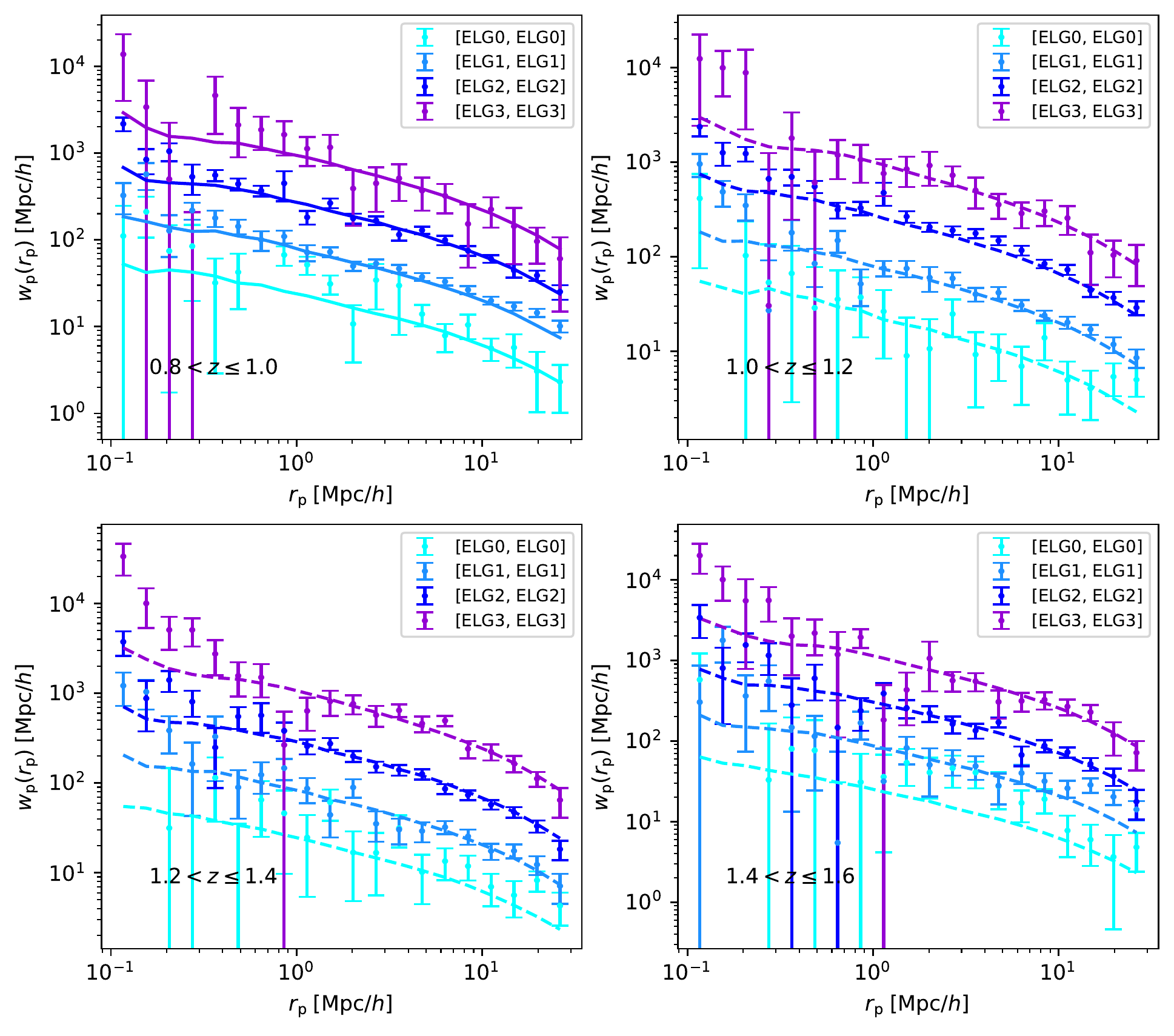}
	\caption{Evolution of projected auto correlation functions for the One-Percent survey ELGs. Different panels represent different redshift intervals. The data points with error bars show the measurements of $\boldsymbol{w}^{\mathrm{obs}}_{\mathrm{p},\mathrm{ELG}}$ for ELG subsamples. Except for the solid lines in the first panel, which are the best-fit results at $0.8<z\leq 1.0$, all dashed lines in the other panels are the predictions by the model. To make a clear presentation, each $\boldsymbol{w}_{\mathrm{p},\mathrm{ELG}}$ has been multiplied by a factor of $3^n$, where $n$ is taken as 0, 1, 2 and 3 from the bottom to the top one.
		\label{fig:wp_evo}}
\end{figure*}

\section{Fitting process and results} \label{sec:results}
In this Section, we fit our model to the projected correlation functions and galaxy number densities. We comprehensively compare the performance of different $P_{\mathrm{sat}}$ models. Using the best-fit parameters, we further predict the correlation functions in redshift-space. We also check if our model is extendable to higher redshifts.

\subsection{A constant $P_{\mathrm{sat}}$ model} \label{sec:psat_constant}
Given the current measurements, we need to check whether a constant $P_{\mathrm{sat}}$ model is sufficient to fit the observations that involve a total of 3 LRG auto correlation functions $\boldsymbol{w}^{\mathrm{obs}}_{\mathrm{p},\mathrm{LRG}}$, 4 ELG auto correlation functions $\boldsymbol{w}^{\mathrm{obs}}_{\mathrm{p},\mathrm{ELG}}$, 16 LRGxELG cross correlation functions $\boldsymbol{w}^{\mathrm{obs}}_{\mathrm{p},\mathrm{LRGxELG}}$ and 3 number densities $n^{\mathrm{obs}}_{\mathrm{LRG}}$ of the LRG subsamples. As we will show shortly, we find that the constant $P_{\mathrm{sat}}$ model is difficult to fit all the correlation functions simultaneously. Good fits to the ELG auto and LRGxELG cross correlations require very different values of $P_{\mathrm{sat}}$. Therefore, we will discuss the two cases separately.

\subsubsection{Fitting without cross correlations} \label{sec:psat_auto}
To constrain our model parameters, we first use the LRG and ELG auto correlations $\boldsymbol{w}^{\mathrm{obs}}_{\mathrm{p},\mathrm{LRG}}$ and $\boldsymbol{w}^{\mathrm{obs}}_{\mathrm{p},\mathrm{ELG}}$, and the LRG number densities $n^{\mathrm{obs}}_{\mathrm{LRG}}$. For subsample $x$ and $y$, we can write their $\chi^{2}_{xy}$ as
\begin{equation}
\chi^2_{xy}=\sum_{k=1}^{N_{r_{\mathrm{p}}}}\sum_{l=1}^{N_{r_{\mathrm{p}}}}\left(w^{\mathrm{obs}}_{\mathrm{p},xy,k}-w^{\mathrm{mod}}_{\mathrm{p},xy,k}\right)C^{-1}_{xy,kl}\left(w^{\mathrm{obs}}_{\mathrm{p},xy,l}-w^{\mathrm{mod}}_{\mathrm{p},xy,l}\right), \label{equ:chi2_auto}
\end{equation}
where $\boldsymbol{w}^{\mathrm{obs}}_{\mathrm{p},xy}$ is the observed correlation function and $\boldsymbol{w}^{\mathrm{mod}}_{\mathrm{p},xy}$ is the model prediction. The inverse of the covariance matrix $\boldsymbol{C}^{-1}_{xy}$ has been multiplied by a bias-correction factor \citep{2007A&A...464..399H}. The total $\chi^{2}$ can be computed as 
\begin{equation}
\begin{aligned}
\chi^2=\sum_{x}\chi^2_{xx} + \sum_{y}\chi^2_{yy} +\sum_{t}\frac{\left(n^{\mathrm{obs}}_{t}-n^{\mathrm{mod}}_{t}\right)^{2}}{\sigma^{2}_{n_t}}, \label{equ:chi2_all_auto}
\end{aligned}
\end{equation}
where $x \in $ [LRG0, LRG1, LRG2], $y \in $ [ELG0, ELG1, ELG2, ELG3] and $t \in$ [LRG1, LRG2, LRG3]. The $n^{\mathrm{obs}}$ and $n^{\mathrm{mod}}$ are the observed and modeled number densities for the LRG subsamples respectively, and the observational uncertainty $\sigma_{n}$ is estimated from the field-to-field variations of 100 jackknife fields. We explore the posterior probability distribution of the model parameters based on Bayesian theory, in which the logarithmic likelihood function is proportional to $-0.5 \chi^{2}$. The priors for the six parameters are set as $10<\log M_0<13$, $0.1<\alpha<0.5$, $1<\beta<5$, $9<\log k<12$, $0<\sigma<1$ and $0<P_{\mathrm{sat}}<1$. Using the code {\tt\string emcee} \citep{2013PASP..125..306F} for Markov Chain Monte Carlo (MCMC) analysis, we run 72 chains each with 1000 steps to sample the entire parameter space. The first 10\% steps of each chain are discarded as burn-in.

The best-fit models for $\boldsymbol{w}^{\mathrm{mod}}_{\mathrm{p},\mathrm{LRG}}$ and $\boldsymbol{w}^{\mathrm{mod}}_{\mathrm{p},\mathrm{ELG}}$ are shown in Figure \ref{fig:wp_auto} as solid lines with distinct colors. The modeled SMF is displayed in the last panel of Figure \ref{fig:wp_auto}. We also present the model predictions for $\boldsymbol{w}^{\mathrm{mod}}_{\mathrm{p},\mathrm{LRGxELG}}$ as dashed lines. Both $\boldsymbol{w}^{\mathrm{obs}}_{\mathrm{p},\mathrm{LRG}}$ and $\boldsymbol{w}^{\mathrm{obs}}_{\mathrm{p},\mathrm{ELG}}$ at $r_{\mathrm{p}}>0.1$ $\mathrm{Mpc}\,h^{-1}$ are well fitted by this model. The posterior distributions of the model parameters are presented in Figure \ref{fig:corner_auto} of Appendix \ref{sec:posterior}. The best-fit parameters are shown in the first column of Table \ref{tab:best-fit}. We can notice that the constraints on the slope $\beta$ is poor. This implies that it is hard to break the parameter degeneracy of SHMR and $P_{\mathrm{sat}}$ using only the $\boldsymbol{w}^{\mathrm{obs}}_{\mathrm{p},\mathrm{ELG}}$. Moreover, due to the significantly high amplitude of $\boldsymbol{w}^{\mathrm{obs}}_{\mathrm{p},\mathrm{ELG}}$ on small scale, the best-fit $P_{\mathrm{sat}}$ parameter tends to be 1. As for the LRGxELG cross correlations, large $P_{\mathrm{sat}}$ makes the model systematically overestimate the observed $\boldsymbol{w}^{\mathrm{obs}}_{\mathrm{p},\mathrm{LRGxELG}}$ on small scales.

\subsubsection{Fitting without ELG auto correlations} \label{sec:psat_cross}
Instead of using $\boldsymbol{w}^{\mathrm{obs}}_{\mathrm{p},\mathrm{ELG}}$, we take the cross correlations $\boldsymbol{w}^{\mathrm{obs}}_{\mathrm{p},\mathrm{LRGxELG}}$ to fit the model. Similarly, we can write the total $\chi^{2}$ as 
\begin{equation}
\begin{aligned}
\chi^2=\sum_{x}\chi^2_{xx} + \sum_{x}\sum_{y}\chi^2_{xy} +\sum_{t}\frac{\left(n^{\mathrm{obs}}_{t}-n^{\mathrm{mod}}_{t}\right)^{2}}{\sigma^{2}_{n_t}}, \label{equ:chi2_all_cross}
\end{aligned}
\end{equation}
where $x \in $ [LRG0, LRG1, LRG2], $y \in $ [ELG0, ELG1, ELG2, ELG3] and $t \in$ [LRG1, LRG2, LRG3]. The best-fit results and the parameter constraints are presented in Figure \ref{fig:wp_cross} (see also Figure \ref{fig:corner_cross} of Appendix \ref{sec:posterior}) and the second column of Table \ref{tab:best-fit}. The model provides a suitable fit for the $\boldsymbol{w}^{\mathrm{obs}}_{\mathrm{p},\mathrm{LRG}}$ and the $\boldsymbol{w}^{\mathrm{obs}}_{\mathrm{p},\mathrm{LRGxELG}}$ and results in better constraints on the parameters of the SHMR model. However, the value of $P_{\mathrm{sat}}$ is only 0.15 in this case. The predicted $\boldsymbol{w}^{\mathrm{mod}}_{\mathrm{p},\mathrm{ELG}}$ are obviously underestimated at $r_{\mathrm{p}}<1$ $\mathrm{Mpc}\,h^{-1}$.

\subsection{A halo mass-dependent $P_{\mathrm{sat}}$ model} \label{sec:psat_model}
As argued in Section \ref{sec:psat_constant}, the best fitting with cross or auto correlations of the LRGs and ELGs can yield very different values of $P_{\mathrm{sat}}$. The lower amplitude of the one-halo term of $\boldsymbol{w}^{\mathrm{obs}}_{\mathrm{p},\mathrm{LRGxELG}}$ implies that there are few ELG satellite galaxies around the massive LRGs. This is consistent with many studies suggesting that the quenched fraction of satellites around massive central galaxies is high \citep{2021MNRAS.500.4004D, PACV}. However, the situation is reversed for the $\boldsymbol{w}^{\mathrm{obs}}_{\mathrm{p},\mathrm{ELG}}$ where the significantly higher clustering signal indicates that there should be more galaxy pairs at $r_{\mathrm{p}}<1$ $\mathrm{Mpc}\,h^{-1}$. Actually, this feature is also reflected in Paper I, although the measurements there had larger uncertainties. In Figure 7 of Paper I, a constant satellite fraction is sufficient to reproduce the auto correlations of the first two ELG subsamples with moderate \oii luminosity. But for the two brightest ELG subsamples whose properties are closer to the DESI samples, the model underestimates the auto correlation on small scales. Since the sample size of VIPERS is small, Paper I was unable to perform more careful analyses. It is worth mentioning that the angular correlations of \oii emitters at $z>1$ measured by \cite{2021PASJ...73.1186O} also show a sudden increase at small scales.

In order to maintain a sufficient number of satellite ELGs in low-mass halos while reasonably considering the quenching effect of satellite galaxies in massive halos, we propose a halo mass-dependent $P_{\mathrm{sat}}$ model as 
\begin{equation}
\begin{aligned}
P_{\mathrm{sat}}(M_{\mathrm{h}})&=\frac{a}{2} \times  \left[1 - \mathrm{erf}(\log M_{\mathrm{h}}-b)\right] \\
&+ \frac{c}{2} \times  \left[ 1-\mathrm{erf}(b-\log M_{\mathrm{h}})\right], \label{equ:p_sat}
\end{aligned}
\end{equation}          
where $\mathrm{erf}$ is the error function and $M_{\mathrm{h}}$ is the host halo mass of satellite galaxies. In this model, $P_{\mathrm{sat}}$ tends to be a constant $a$ at $\log M_{\mathrm{h}} < b$, while decreases as the halo mass increases at $\log M_{\mathrm{h}} > b$ and finally reaches a constant $c$. As shown in Section \ref{sec:psat_auto} (see also Figure \ref{fig:corner_auto}), due to the high amplitude of $\boldsymbol{w}^{\mathrm{obs}}_{\mathrm{p},\mathrm{ELG}}$ at small scales, the best-fit value of $P_{\mathrm{sat}}$ converges to 1 for the ELG satellite galaxies in low-mass halos. We therefore fix the parameter $a = 1$ in Equation \ref{equ:p_sat}.

Combining all the measurements, the total $\chi^{2}$ can be computed as 
\begin{equation}
\chi^2= \sum_{x}\chi^2_{xx} + \sum_{y}\chi^2_{yy} + \sum_{x}\sum_{y}\chi^2_{xy} + \sum_{t}\frac{\left(n^{\mathrm{obs}}_{t}-n^{\mathrm{mod}}_{t}\right)^{2}}{\sigma^{2}_{n_t}}, \label{equ:chi2_all}
\end{equation}
where $x \in $  [LRG0, LRG1, LRG2], $y \in $ [ELG0, ELG1, ELG2, ELG3] and $t \in$ [LRG1, LRG2, LRG3]. The fitting results for this model are displayed in Figure \ref{fig:wp}, Figure \ref{fig:corner} of Appendix \ref{sec:posterior} and the third column of Table \ref{tab:best-fit}. We present all the best-fit SHMR and $P_{\mathrm{sat}}$ models in Figure \ref{fig:shmr}.

In general, the reduced $\chi^2/\mathrm{dof} = 0.89$ indicates that the overall fit is reasonable. The model can reproduce both the $\boldsymbol{w}^{\mathrm{obs}}_{\mathrm{p},\mathrm{LRGxELG}}$ at $r_{\mathrm{p}}>0.1$ $\mathrm{Mpc}\,h^{-1}$ and the $\boldsymbol{w}^{\mathrm{obs}}_{\mathrm{p},\mathrm{ELG}}$ at $r_{\mathrm{p}}>0.3$ $\mathrm{Mpc}\,h^{-1}$, and thus overcomes the shortcomings of the constant $P_{\mathrm{sat}}$ model. The five parameters of the unified SHMR model can be well constrained. This illustrates that the combination of massive LRG samples and low-mass ELG samples can effectively help us to determine the SHMR in a wide range of stellar mass. 

In the halo mass-dependent $P_{\mathrm{sat}}$ model, the parameter $a$, which represents ELG satellite probability in small halos, is set to 1. However, such a $P_{\mathrm{sat}}$ value is still difficult to reproduce the one-halo term of $\boldsymbol{w}^{\mathrm{obs}}_{\mathrm{p},\mathrm{ELG}}$ at $r_{\mathrm{p}}<0.3$ $\mathrm{Mpc}\,h^{-1}$, which is still higher. One possible explanation is 1-halo galaxy conformity \citep[e.g.,][]{2006MNRAS.366....2W, 2018MNRAS.480.2031C, 2022MNRAS.511.1789Z, 2022arXiv221010068H}, which describes a phenomenon that the physical properties of central and satellite galaxies in the same halo are correlated. There is still a debate about whether galaxy conformity exists. By comparing the SDSS observations to mock catalogs, \cite{2018MNRAS.480.2031C} argued that the 1-halo conformity is not real and could be caused by group-finding systematics. Using the abundance and weak lensing measurements of SDSS clusters, \cite{2022MNRAS.511.1789Z} detected a halo mass-dependent galaxy conformity between the stellar mass of bright central galaxies (BCGs) and the cluster satellite richness. \cite{2022arXiv221010068H} have shown that if a halo already contains an ELG satellite, its central galaxy is twice as likely to be an ELG. They further introduce a free parameter in the HOD to interpret this conformity effect. Recently, \cite{abacusELG_Rocher} also introduce a parameterized conformity bias in the HOD model to reproduce the auto correlation of ELGs in the One-Percent survey. Nonetheless, the conformity of ELGs may have more complicated dependencies on different physical properties such as stellar mass, emission line strength and environment. In our future work, we will develop a comprehensive model to quantify the conformity of ELGs.

\subsection{Checking the correlation functions in redshift-space} \label{sec:redshift space}
Using the best-fit SHMR and $P_{\mathrm{sat}}$ models, we make a prediction for the multiple moments $\xi_0(s)$, $\xi_2(s)$ and $\xi_4(s)$ in redshift-space. In order to make a fair comparison with the observations, we have incorporated redshift measurement errors and galaxy velocity bias in our model. With multiple independent spectroscopic observations for the same object, \cite{2023ApJ...943...68L} have presented the uncertainty distributions of the redshifts for BGSs, LRGs and ELGs in DESI. Accordingly, we assume that the redshift uncertainty follows a Gaussian distribution with a scatter of $\sigma_z$, where $\sigma_z$ is fixed to $40$ $\mathrm{km\,s^{-1}}$ and $10$ $\mathrm{km\,s^{-1}}$ for the modeled LRGs and ELGs respectively. Considering that a central galaxy is typically not at rest with respect to its host halo \citep{2003ApJ...590..654Y}, we also randomly assign a Gaussian distribution with $\sigma_{\mathrm{c}} = \alpha_{\mathrm{c}} \times \sigma_v$ for each halo to account for the 1-D velocity bias. Here $\alpha_{\mathrm{c}}$ is set as 0.22 that is determined by \cite{2015MNRAS.446..578G} using BOSS CMASS galaxies, while $\sigma_v = (GM_{\mathrm{vir}}/(2R_{\mathrm{vir}}))^{0.5}$ is the 1-D velocity dispersion of the halo.

The observed multiple moments and our model predictions are displayed in Figure \ref{fig:xi0}, \ref{fig:xi2} and \ref{fig:xi4}. Here we omit the $\xi_2(s)$ and $\xi_4(s)$ of the auto correlations of LRG2 due to their too noisy measurements. We show that although we use only $\boldsymbol{w}^{\mathrm{obs}}_{\mathrm{p}}$ in fitting, our model can well reproduce the auto correlations of LRGs and the cross correlations of LRGsxELGs in redshift-space. Similar to the case in real-space, for the auto correlations of ELGs, the predicted $\xi_0(s)$, $\xi_2(s)$ and $\xi_4(s)$ are underestimated on small scales. Nevertheless, at $s > 1$ $\mathrm{Mpc}\,h^{-1}$ where the Kaiser effect plays a dominant role, our model is sufficient to reproduce the observed multiple moments.

\subsection{Extending the ELG-halo connection to higher redshifts} \label{sec:redshift evolution}
Since the overlap between LRGs and ELGs is only in the redshift range of $0.8<z<1.0$, beyond which we cannot constrain the ELG-halo connection by their cross correlations. We need to check whether our model can be safely extended to higher redshifts.

We divide all the ELG samples within $0.8<z\leq1.6$ into four redshift bins and measure their auto correlations. However, for redshift $0.6<z \leq 0.8$, due to the limited number of ELG samples and the incompleteness at the bright end (see also Figure \ref{fig:oii}), we do not consider this redshift range here. Using four individual snapshots at redshifts $0.92$, $1.09$, $1.27$ and $1.47$, we calculate the modeled $\boldsymbol{w}^{\mathrm{mod}}_{\mathrm{p}}$ for each redshift bin using the best-fit parameters presented in Section \ref{sec:psat_model}. We exhibit the evolution of $\boldsymbol{w}^{\mathrm{obs}}_{\mathrm{p},\mathrm{ELG}}$ as well as the model predictions in Figure \ref{fig:wp_evo}. At redshift $1.0<z\leq1.6$, the model predictions can match well the measurements at $r_{\mathrm{p}}>0.5$ $\mathrm{Mpc}\,h^{-1}$. This implies that the SHMR and the ELG-halo connection evolve weakly from redshift $z=0.8$ to $1.6$. Although our model is obtained at $z\sim 0.9$ and has only seven free parameters, it can still reproduce the observations over the entire range of $0.8<z<1.6$ given the current measurement errors. This will allow us to build an ELG mock catalog for the DESI One-Percent and Y1 surveys.

\begin{figure*}
	\centering
	\includegraphics[scale=0.8]{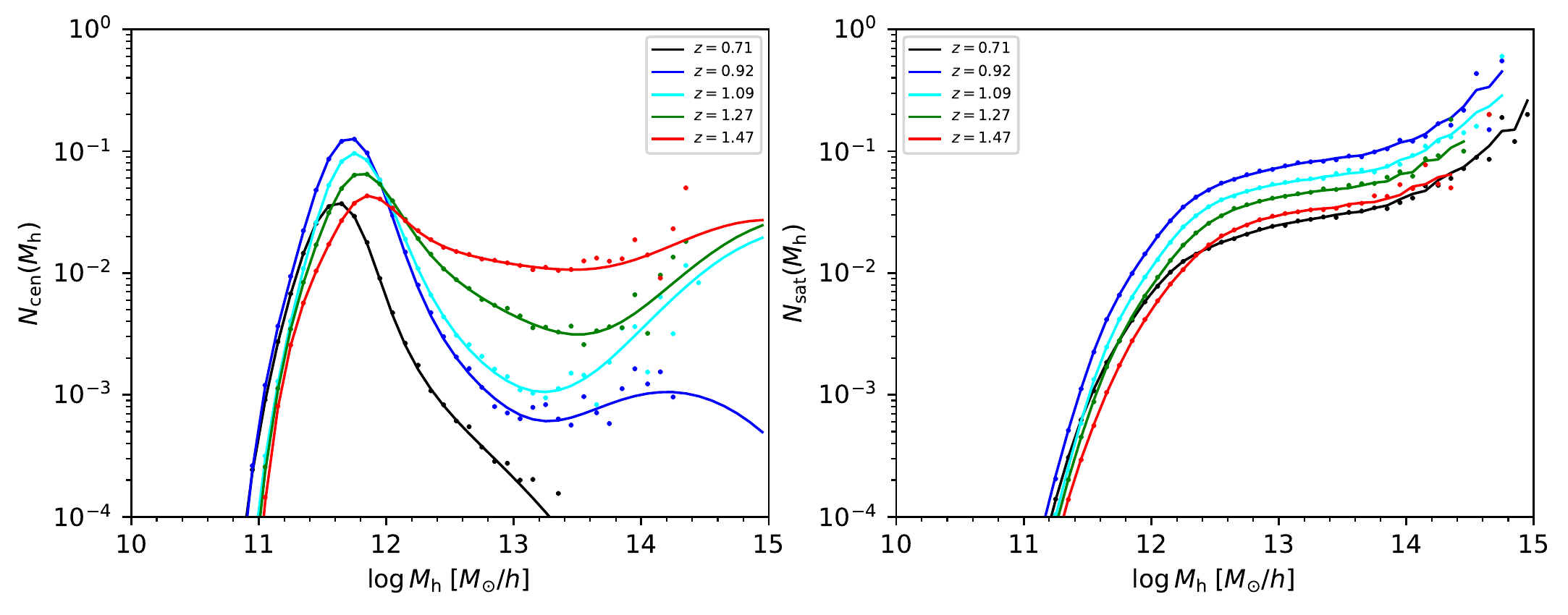}
	\caption{HODs of the ELGs in the One-Percent survey. The central occupation $N_{\rm{cen}}(M_{\rm{h}})$ and the satellite occupation $N_{\rm{sat}}(M_{\rm{h}})$ are presented in the left and right panels respectively. The solid lines denote the results of the theoretical calculations using Equation \ref{equ:hod}. The colored dots show the averaged HODs measured directly from 10 random realizations. The different colors correspond to the HODs at different redshifts.
		\label{fig:hod}}
\end{figure*}

\begin{figure*}
	\centering
	\includegraphics[scale=0.6]{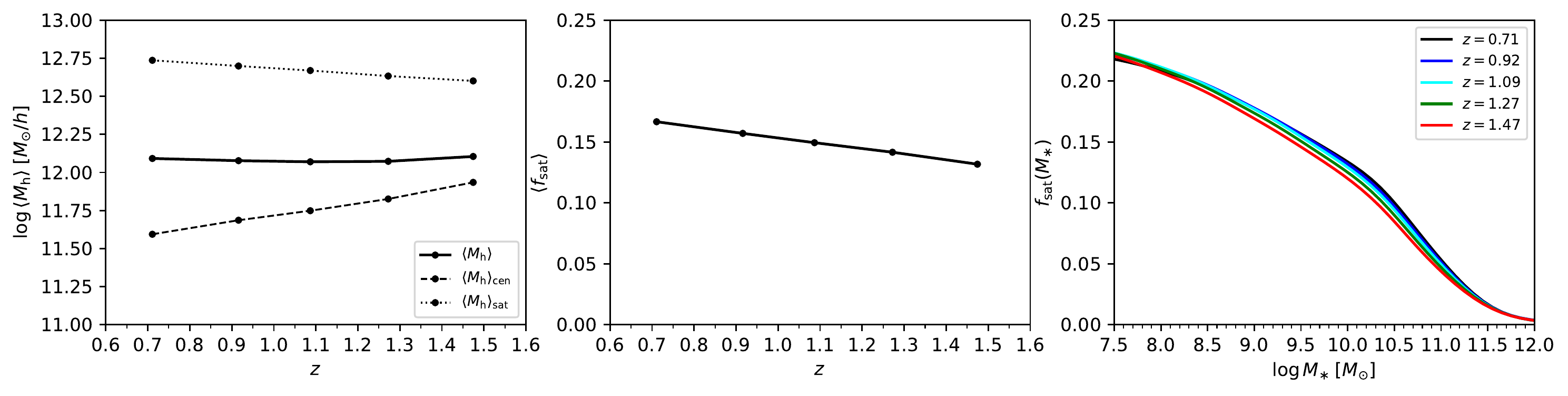}
	\caption{Evolution of the mean host halo mass and the satellite fraction of the ELGs in our model predictions. The left panel shows the mean host halo mass for all ELGs $\left \langle  M_{\mathrm{h}} \right \rangle $ (solid line), central ELGs $\left \langle  M_{\mathrm{h}} \right \rangle_\mathrm{cen}$ (dashed line) and satellite ELGs $\left \langle  M_{\mathrm{h}} \right \rangle _\mathrm{sat}$ (dotted line). The middle panel shows the evolution of the mean ELG satellite fraction $\left \langle f_{\mathrm{sat}}\right \rangle$. The right panel presents the ELG satellite fraction $ f_{\mathrm{sat}}(M_{\ast})$ as a function of stellar mass at each redshift. 
		\label{fig:elg_mean_halo_mass_satellite_fraction}}
\end{figure*}

\section{HOD of ELGs} \label{sec:hod_of_ELG}
\subsection{Theoretical derivation} 
\label{sec:hod}
One of the advantages of our abundance matching approach is that we can theoretically derive the HOD of the ELGs without assuming a complicated parameterized HOD model.

Given a halo with mass $M_{\mathrm{h}}$, the probability of its central galaxy becoming an ELG can be calculated as
\begin{equation}
P^{\rm{ELG}}(M_{\rm{h}})= \int^{+ \infty}_{-\infty}F^{\rm{ELG}}(M_{\ast})p(M_{\ast}|M_{\rm{h}})\, \mathrm{d}\log M_{\ast}, \label{equ:p_elg(M_h)}
\end{equation}

where $p(M_{\ast}|M_{\rm{h}})$ is the SHMR in Equation \ref{equ:p(M_star|M_h)}, and $F^{\rm{ELG}}(M_{\ast}) = \bar{n}^{\mathrm{obs}}\left(M_{\ast}\right)/\bar{n}^{\mathrm{can}}\left(M_{\ast}\right)$ is the observed ELG fraction in the modeled ELG candidates as defined in Equation \ref{equ:f(M_star)}. Similarly, given a subhalo with mass $M_{\mathrm{s}}$, the probability of the satellite galaxy in it becoming an ELG can be written as
\begin{equation}
\begin{aligned}
P^{\rm{ELG}}&(M_{\rm{s}},M_{\rm{h}})\\
&= \int^{+ \infty}_{-\infty}F^{\rm{ELG}}(M_{\ast})p(M_{\ast}|M_{\rm{s}})P_{\rm{sat}}(M_{\rm{h}})\, \mathrm{d}\log M_{\ast}, \label{equ:p_elg(M_s)}
\end{aligned}
\end{equation}
where we select the halo mass-dependent $P_{\rm{sat}}(M_{\rm{h}})$ model shown in Equation \ref{equ:p_sat} as the $P_{\rm{sat}}(M_{\rm{h}})$.

The key step is to compute the number density of ELG candidates in the model and derive $F^{\rm{ELG}}(M_{\ast})$. Firstly, the total number density $\bar{n}^{\rm{can}}(M_{\ast})$ of the ELG candidates can be written as
\begin{equation}
\bar{n}^{\rm{can}}(M_{\ast}) = \bar{n}^{\rm{can}}_{\rm{cen}}(M_{\ast}) + \bar{n}^{\rm{can}}_{\rm{sat}}(M_{\ast}),
\end{equation}
where $\bar{n}^{\rm{can}}_{\rm{cen}}(M_{\ast})$ and $\bar{n}^{\rm{can}}_{\rm{sat}}(M_{\ast})$ are the number densities of the central and satellite ELG candidates, respectively. Then, $\bar{n}^{\rm{can}}_{\rm{cen}}(M_{\ast})$ can be calculated via
\begin{equation}
\begin{aligned}
\bar{n}^{\rm{can}}_{\rm{cen}}(M_{\ast})&=\int^{+\infty}_{-\infty}P(M_{\ast}|M_{\rm{h}})n(M_{\rm{h}}) \,  \mathrm{d} \log M_{\mathrm{h}} \\
&= \sum_{i} P(M_{\ast}|M_{\mathrm{h},i})\bar{n}(M_{\rm{h},i}),
\end{aligned}
\end{equation}
where $P(M_{\ast}|M_{\rm{h}})$ is the probability that a halo hosts a central galaxy with stellar masses between $\log M_{\ast}-\Delta \log M_{\ast}/2$ and $\log M_{\ast}+\Delta \log M_{\ast}/2$, and $n(M_{\rm{h}})$ is the halo mass function. To numerically evaluate this integration, from $10^{10}$ to $10^{15} \, M_{\odot}\,h^{-1}$, we divide the halo samples into 50 tiny bins in logarithmic space with a width of $\Delta \log M_{\rm{h}} = 0.1$. For the $i$-th bin, we can calculate the probability $P(M_{\ast}|M_{\mathrm{h},i})$ by
\begin{equation}
P(M_{\ast}|M_{\mathrm{h},i}) 
= \int^{\log M_{\ast}+\Delta \log M_{\ast}/2}_{\log M_{\ast}-\Delta \log M_{\ast}/2} p(M_{\ast}|M_{\mathrm{h},i})\, \mathrm{d} \log M_{\ast},
\end{equation}
and measure $\bar{n}(M_{\rm{h},i})$ directly from simulation. Analogously, the number density of satellite candidates $\bar{n}^{\rm{can}}_{\rm{sat}}(M_{\ast})$ can also be derived in this way
\begin{equation}
\begin{aligned}
\bar{n}^{\rm{can}}_{\rm{sat}}(M_{\ast})& \\ =\iint^{+\infty}_{-\infty}&P(M_{\ast}|M_{\rm{s}}) P_{\rm{sat}}(M_{\rm{h}}) \\
& \times n_{\rm{s}}(M_{\rm{s}}|M_{\rm{h}})n(M_{\rm{h}}) \,  \mathrm{d} \log M_{\mathrm{s}}  \mathrm{d} \log M_{\mathrm{h}} \\
=\sum_i \sum_j& P(M_{\ast}|M_{\mathrm{s},j}) P_{\rm{sat}}(M_{\mathrm{h},i})\bar{n}_{\rm{s}}(M_{\mathrm{s},j}|M_{\mathrm{h},i})\bar{n}(M_{\mathrm{h},i}),
\end{aligned}
\end{equation}
where subhalo mass function $\bar{n}_{\rm{s}}(M_{\mathrm{s},j}|M_{\mathrm{h},i})$ is measured in $50 \times 50$ grids, and the probability $P(M_{\ast}|M_{\mathrm{s},j})$ for the $j$-th subhalo mass bin can be computed via
\begin{equation}
P(M_{\ast}|M_{\mathrm{h},j}) 
= \int^{\log M_{\ast}+\Delta \log M_{\ast}/2}_{\log M_{\ast}-\Delta \log M_{\ast}/2} p(M_{\ast}|M_{\mathrm{h},j})\, \mathrm{d} \log M_{\ast},
\end{equation}
Finally, the HOD forms of ELGs can be expressed as follows
\begin{equation}
\begin{aligned}
N_{\rm{cen}}(M_{\rm{h}})&= P^{\rm{ELG}}(M_{\rm{h}}) \\
N_{\rm{sat}}(M_{\rm{h}})&=\int^{+\infty}_{-\infty}P^{\rm{ELG}}(M_{\rm{s}},M_{\rm{h}})n_{\rm{s}}(M_{\rm{s}}|M_{\rm{h}})\,  \mathrm{d} \log M_{\mathrm{s}} \\
&=\sum_i P^{\rm{ELG}}(M_{\mathrm{s},i}|M_{\rm{h}})\bar{n}_{\rm{s}}(M_{\mathrm{s},i}|M_{\rm{h}}). \label{equ:hod}
\end{aligned}
\end{equation}
Using the above formula, we can obtain the central and satellite occupation numbers of ELGs for a given halo with mass $M_{\rm{h}}$. 

In Figure \ref{fig:hod}, we present the theoretical HODs calculated by Equation \ref{equ:hod} as solid lines. To further make a self-consistent test, we perform 10 random realizations to populate ELGs in simulations and directly measure their HODs. The average results from the 10 realizations are shown as dots in Figure \ref{fig:hod}. 

The central occupations of our model can clearly decompose ELGs into two different populations. On the one hand, at the low-mass end, $N_{\rm{cen}}(M_{\rm{h}})$ shows a similar peak at $10^{11.5}$ to $10^{12} \, M_{\odot}\,h^{-1}$, which is consistent with other studies \citep[e.g.,][]{2016MNRAS.461.3421F, 2019ApJ...871..147G, 2021MNRAS.502.3599H, 2021PASJ...73.1186O}. As the redshift increases, the location of the peak gradually shifts towards the high-mass end. It may be due to the fact that the selected ELG samples in DESI have relatively higher stellar mass at higher redshift (see also the right panel of Figure \ref{fig:lrg_and_elg_number_density}) and hence tend to reside in larger halos. Beyond the peaks, $N_{\rm{cen}}(M_{\rm{h}})$ begins to decay from $10^{12}$ to $10^{13} \, M_{\odot}\,h^{-1}$, which reflects the fact that star formation in massive galaxies has almost been stopped \citep{2020ApJ...895..100X}. However, on the other hand, there is an upturn in $N_{\rm{cen}}(M_{\rm{h}})$ at about  $M_{\rm{h}} > 10^{13.5} \, M_{\odot}\,h^{-1}$. This may actually correspond to another population of ELGs. Central galaxies in massive halos are more likely to host AGN, and the high-energy radiation from their central engine can ionize the surrounding gas and produce strong emission lines \citep{2013MNRAS.428.1498C}. Therefore, although low-mass star-forming ELGs dominate our current sample, we also need to pay more attention to those massive ELGs that may be related to AGN activities.

For the satellite occupation $N_{\rm{sat}}(M_{\rm{h}})$, its shape is mainly determined by the $P_{\rm{sat}}(M_{\rm{h}})$ model. At $M_{\rm{h}} < 10^{12} \, M_{\odot}\,h^{-1}$ where $P_{\rm{sat}}$ is close to 1, $N_{\rm{sat}}$ increases rapidly with the host halo mass in a power-law form. Then, as $P_{\rm{sat}}$ decreases with $M_{\rm{h}}$ at $10^{12} < M_{\rm{h}} < 10^{14} \, M_{\odot}\,h^{-1}$, $N_{\rm{sat}}$ keeps almost a constant because the increase in the number of subhalos with $M_{\rm{h}}$ roughly cancels out the decrease of $P_{\rm{sat}}$. Above $ M_{\rm{h}} > 10^{14} \, M_{\odot}\,h^{-1}$ where $P_{\rm{sat}}$ stays at a constant about 0.04,  $N_{\rm{sat}}$ is nearly proportional to the host halo mass $M_{\rm{h}}$, as the number of subhalos within a host halo is approximately proportional to $M_{\rm{h}}$ \citep[e.g.,][]{2004MNRAS.355..819G,2012MNRAS.425.2169G,2008MNRAS.386.2135G,2016MNRAS.457.1208H}. 

From the derived HOD, we find that the HOD of ELGs depends on the host halo mass and the observed redshift in a complicated way. It would be difficult to find an analytical form to represent the HOD of ELGs at different redshifts. And the analytical models at different redshifts may have different sets of parameters. It is worth mentioning that although our model is obtained from the SHAM method, our derived HOD forms can still be applied to populating ELGs with halos in low-resolution simulations where subhalos are not resolved. Therefore, we believe that the SHAM approach presented here has advantages over the conventional HOD method, and can be easily applied to N-body simulations to produce mock catalogs for ELGs.

\begin{deluxetable}{ccccc}
	\tablenum{4}
	\tablecaption{The values of mean halo mass and satellite fraction of ELGs in our model. }
	\label{tab:elg_mean_halo_mass_satellite_fraction}
	\tablehead{  Redshift &
		\colhead{$\log \left \langle  M_{\mathrm{h}} \right \rangle  $} & \colhead{$\log \left \langle  M_{\mathrm{h}} \right \rangle_\mathrm{cen} $ } &\colhead{$\log \left \langle  M_{\mathrm{h}} \right \rangle_\mathrm{sat} $}  &\colhead{$\left \langle f_{\mathrm{sat}}\right \rangle$}
	}
	\startdata
	$z=0.71$ & $12.091$ & $11.594$ & $12.736$ & $0.167$ \\
	$z=0.92$ & $12.077$ & $11.686$ & $12.699$ & $0.157$ \\
	$z=1.09$ & $12.070$ & $11.748$ & $12.669$ & $0.149$ \\
	$z=1.27$ & $12.073$ & $11.825$ & $12.633$ & $0.142$ \\
	$z=1.47$ & $12.104$ & $11.934$ & $12.601$ & $0.132$ \\
	\enddata
	\tablecomments{The unit of $\left \langle  M_{\mathrm{h}} \right \rangle$, $ \left \langle  M_{\mathrm{h}} \right \rangle_\mathrm{cen} $ and $ \left \langle  M_{\mathrm{h}} \right \rangle_\mathrm{sat} $ is $M_{\odot}\,h^{-1}$.}
\end{deluxetable}

\subsection{Mean host halo mass and satellite fraction of ELGs in the model}\label{sec:mean_halo_mass}
For comparison with other HOD studies, we calculate the mean host halo mass $\left \langle M_{\rm{h}} \right \rangle$ of ELGs and the ELG satellite fraction $\left \langle f_{\mathrm{sat}}\right \rangle$ in the model. Based on the derived HODs at different redshifts, we can calculate the average mass $\left \langle M_{\mathrm{h}} \right \rangle$ of the host halos for ELGs:

\begin{equation}
\begin{aligned}
\left \langle M_{\mathrm{h}} \right \rangle &= \frac{\int^{+\infty}_{-\infty} N_{\mathrm{ELG}}(M_{\mathrm{h}})n(M_{\mathrm{h}}) M_{\mathrm{h}} \,\mathrm{d} \log M_{\mathrm{h}} }{\int^{+\infty}_{-\infty} N_{\mathrm{ELG}}(M_{\mathrm{h}})n(M_{\mathrm{h}}) \mathrm{d} \log M_{\mathrm{h}}} \\
&= \frac{\sum_i N_{\mathrm{ELG}}(M_{\mathrm{h},i}) \bar{n}(M_{\mathrm{h},i})M_{\mathrm{h},i}}{\sum_i N_{\mathrm{ELG}}(M_{\mathrm{h},i}) \bar{n}(M_{\mathrm{h},i})}, 
\end{aligned}
\end{equation}
where $N_{\mathrm{ELG}}(M_{\mathrm{h}}) = N_{\mathrm{cen}}(M_{\mathrm{h}}) + N_{\mathrm{sat}}(M_{\mathrm{h}})$ is the total occupation number of ELGs. We present the mean halo mass $\left \langle M_{\mathrm{h}} \right \rangle$ as function of redshifts in the left panel of Figure \ref{fig:elg_mean_halo_mass_satellite_fraction}. Similarly, we also calculate the mean halo mass for central ELGs $\left \langle  M_{\mathrm{h}} \right \rangle_\mathrm{cen}$ and satellite ELGs $\left \langle  M_{\mathrm{h}} \right \rangle_\mathrm{sat}$, and show them in Figure \ref{fig:elg_mean_halo_mass_satellite_fraction}. We can see that the central ELGs tend to occupy larger halos at higher redshift, while the host halo mass of the satellite ELGs decrease with redshift. But the overall $\left \langle M_{\mathrm{h}} \right \rangle$ is almost a constant across the probed redshift range. \cite{abacusELG_Rocher} also show that the evolution of the host halo mass of ELG with redshift is weak ($\left \langle \log M_{\mathrm{h}} \right \rangle \sim 11.8$ at both $0.8<z<1.1$ and $1.1<z<1.6$), although their values are slightly lower.

We can further compute the mean ELG satellite fraction $\left \langle f_{\mathrm{sat}}\right \rangle$ in the model:
\begin{equation}
\begin{aligned}
\left \langle f_{\mathrm{sat}}\right \rangle &= \frac{\int^{+\infty}_{-\infty} N_{\mathrm{sat}}(M_{\mathrm{h}})n(M_{\mathrm{h}}) \,\mathrm{d} \log M_{\mathrm{h}} }{\int^{+\infty}_{-\infty} N_{\mathrm{ELG}}(M_{\mathrm{h}})n(M_{\mathrm{h}}) \mathrm{d} \log M_{\mathrm{h}}} \\
&= \frac{\sum_i N_{\mathrm{sat}}(M_{\mathrm{h},i}) \bar{n}(M_{\mathrm{h},i})}{\sum_i N_{\mathrm{ELG}}(M_{\mathrm{h},i}) \bar{n}(M_{\mathrm{h},i})}. 
\end{aligned}
\end{equation}

The $\left \langle f_{\mathrm{sat}}\right \rangle$ at each redshift is shown in the middle panel of Figure \ref{fig:elg_mean_halo_mass_satellite_fraction}. It shows a monotonically decreasing evolution with increasing redshift. This trend is similar to the study of \cite{2019ApJ...871..147G} in which the $\left \langle f_{\mathrm{sat}}\right \rangle$ of the eBOSS ELG samples also decreases with redshift. In addition, since the $P_{\rm{sat}}(M_{\rm{h}})$ depends on the halo mass, the ELG satellite fraction in our model is not a constant. Here we also represent $ f_{\mathrm{sat}}(M_{\ast})$ as a function of stellar mass at each redshift, and show them in the right panel of Figure \ref{fig:elg_mean_halo_mass_satellite_fraction}. All the values of $\left \langle M_{\mathrm{h}} \right \rangle$ and $\left \langle f_{\mathrm{sat}}\right \rangle$ are listed in Table \ref{tab:elg_mean_halo_mass_satellite_fraction}. 

\section{Discussion} \label{sec:discussion}
There is a series of works in parallel that use different methods for constructing the galaxy-halo connection for LRGs and ELGs in the One-percent survey \citep[e.g.,][]{abacusLRGQSO_Yuan,LRG_Ereza,abacusELG_Rocher,ELG_Lasker,multi-tracerHOD_Yuan,inclusiveSHAM,overviewSHAM}.

For example, \cite{abacusLRGQSO_Yuan} perform a comprehensive HOD analysis for LRGs and QSOs using {\tt\string AbacusSummit} simulation \citep{2021MNRAS.508.4017M}. They combine the standard HOD model with incompleteness parameter, galaxy velocity bias, and galaxy assembly bias. They test these possible HOD extensions by fitting LRG correlation functions at different redshifts. The best-fit values of the velocity bias are mostly consistent with the previous BOSS results. They demonstrate that the galaxy assembly bias has almost no effect on the fitting given the current measurement precision.

For the HOD modeling of ELGs, \cite{abacusELG_Rocher} explore the performance of four different HOD forms in the fitting. In particular, to recover the strong clustering of ELGs on small scales, they introduce a strict central-satellite conformity bias that only allows satellite pairs to exist in the halos whose central galaxy is an ELG. They also argue that the velocity dispersion of ELG satellites should be larger than that of the dark matter particles.

\cite{inclusiveSHAM} develop a generalized SHAM approach to model the LRGs, ELGs and QSOs in {\tt\string UNIT} simulation \citep{2019MNRAS.487...48C}. Besides the intrinsic dispersion of the galaxy-halo relation, they further model the uncertainty of the redshift measurement and add an incompleteness parameter to account for the absence of galaxies in the massive halos. They modeled the clustering only at large scale $s>5$ $\mathrm{Mpc}\,h^{-1}$. In particular for ELGs, they show that the ELG satellite fraction $f_{\mathrm{sat}}$ in the model should be suppressed to $\sim 4 \%$ to recover the observed redshift clustering.

Using {\tt\string Uchuu} simulations \citep{2021MNRAS.506.4210I}, \cite{overviewSHAM} generate lightcone mocks for four DESI tracers through the SHAM method. For the LRGs, they first adopt a complete SMF from the PRIMUS survey to implement the SHAM process, and then down-sample the galaxies in simulation based on the observed LRG SMF. They find a systematically low clustering of the mocks below $\sim 5$ $\mathrm{Mpc}\,h^{-1}$, probably due to the different properties of the PRIMUS and DESI samples. For the ELGs, they assume that the maximum circular velocity of the host halos (subhalos) is a Gaussian distribution with its amplitude normalized to the observed number density. In their model, the ELG satellite fraction $f_{\mathrm{sat}}$ is considered as a free parameter. The observed ELG clustering can be reproduced by their model at $s>4$ $\mathrm{Mpc}\,h^{-1}$.

Each of the above studies has its pros and cons. In comparison, the main advantages of our method can be summarized as follows. (1) We have used the auto and cross correlations of ELGs and LRGs jointly to simultaneously constrain the SHMR (SMF) and the ELG-halo relation. The former can be used to generate LRG mocks. (2) Because we model ELGs and LRGs jointly, we demonstrate that a mass-dependent $P_{\mathrm{sat}}$ model is essential to describe satellite ELGs in both small and large halos. (3) With our SHAM model, we can derive the conventional HOD  which can also be applied to coarse simulations. (4) Given the current measurement uncertainty, our model for ELGs can be extended to the entire redshift range of $0.8<z<1.6$, without introducing a large number of parameters at each redshift. Our model is accurate in describing the clustering of LRGs and ELGs down to sub $\mathrm{Mpc}\,h^{-1}$ scales. 

\section{Summary} \label{sec:summary}
In this work, we extend our novel method \citep{2022ApJ...928...10G} to accurately construct galaxy-halo connections for LRGs and ELGs using the DESI One-Percent survey. Our method can simultaneously constrain the SHMR for normal galaxies and in particular the ELG-halo connection in the One-Percent survey. We summarize our main results as follows.

\begin{enumerate}
	\item Using the galaxy catalog from DESI One-Percent survey, we perform a SED fitting and measure the apparent SMFs for the LRG and ELG samples. At redshift $0.8<z\leq 1.0$, we divide the LRG and ELG samples into eight subsamples based on their stellar masses. We estimate the LRG auto correlations, ELG auto correlations and LRGxELG cross correlations for all galaxy subsamples.
	\item Combining the abundance matching technique and a high-resolution N-body simulation from {\tt\string CosmicGrowth} \citep{2019SCPMA..6219511J}, we simultaneously model the galaxy clustering for LRGs and ELGs. We adopt the SHMR model proposed by \citet{2010MNRAS.402.1796W} to establish the normal galaxy-halo relation. Given the SHMR, normal galaxies can be populated to halos and subhalos in the simulation. We select stellar mass-complete LRG samples from the massive normal galaxies. We select ELGs in two steps. We first consider all central galaxies as ELG candidates, while reducing the probability that satellite galaxies become ELG candidates with the adjusting parameter $P_{\mathrm{sat}}$. Then, we calculate the ELG fraction $F^{\rm{ELG}}(M_{\ast})$, which is the ratio of the number density of ELGs in the observation to that of the ELG candidates in the model. By assigning a probability $F^{\rm{ELG}}(M_{\ast})$ to each candidate, the ELG samples can be randomly selected from these candidates. 
	\item We utilize MCMC analysis to explore the parameter space of our model. With the LRG samples, the massive end of SHMR can be well determined, while the ELG samples provide much information for the SHMR down to $10^{8.5}$ $M_{\odot}$. We also do a test for the different $P_{\mathrm{sat}}$ models. We find that the ELG auto correlations and the LRGxELG cross correlations can lead to very different $P_{\mathrm{sat}}$ values. 
     Thus, we propose a host halo mass-dependent $P_{\mathrm{sat}}$ model. This model can reasonably reduce the $P_{\mathrm{sat}}$ for the massive halos while retaining a sufficient number of satellite ELGs in small halos. Our model can well reproduce the SMF of LRG at the massive end, the LRG auto correlations, the LRGxELG cross correlations and the ELG auto correlations at $r_{\mathrm{p}}>0.3$ $\mathrm{Mpc}\,h^{-1}$. Using this model, we further predict the multiple moments in redshift-space. Although our model has only seven parameters that are fully fixed by the projected correlations at $z\sim 0.9$, our model predictions are consistent with the redshift-space correlatins at $s>1$ $\mathrm{Mpc}\,h^{-1}$. We also check if our model is valid for other redshifts covered by DESI, and we find that our model can match well the ELG auto correlations at $r_{\mathrm{p}}>0.5$ $\mathrm{Mpc}\,h^{-1}$ for the entire redshift range  $0.8<z\leq 1.6$. Thus our model can be used for generating ELG mock catalogs. 
     	\item Based on our model, we theoretically derive the HOD forms for ELGs in the One-Percent survey at different redshifts. The shape of the central occupations indicates that ELGs can be plainly divided into two populations. Star-forming galaxies dominate the low-mass end of ELGs, which tend to reside in halos with mass $10^{11.5}$ to $10^{13} \, M_{\odot}\,h^{-1}$. We have seen the upturn of the HOD at halo mass $>10^{13} \, M_{\odot}\,h^{-1}$, which might indicate that AGN galaxies may contribute to the massive end of the ELG SMF. Overall, the HODs of both central and satellite galaxies show a complicated dependence on host halo mass and redshift, which may indicate that it is challenging to find simple analytical forms to represent the HOD of ELGs in the DESI survey and/or future ELG surveys. Our model can describe the evolving HODs of ELGs with just seven parameters, without introducing different sets of parameters at different redshifts.
	
\end{enumerate}

In our subsequent work, we will investigate the origin of the strong clustering of ELGs on very small scales (less than a few hundred $\mathrm{kpc}\,h^{-1}$) and further refine our model. Meanwhile, we will also generate realistic ELG lightcone mock catalogs for the DESI One-Percent and Y1 surveys based on our model, and study its impact on future DESI cosmological probes such as BAOs, redshift distortion, and weak lensing.

\section*{Acknowledgments}
We thank Chris Blake, Zheng Zheng and Sandy Yuan for their great help during the DESI Collaboration Wide Review. 

The work is supported by NSFC (12133006, 11890691, 11621303), by grant No. CMS-CSST-2021-A03, and by 111 project No. B20019. We gratefully acknowledge the support of the Key Laboratory for Particle Physics, Astrophysics and Cosmology, Ministry of Education. This work made use of the Gravity Supercomputer at the Department of Astronomy, Shanghai Jiao Tong University.

This material is based upon work supported by the U.S. Department of Energy (DOE), Office of Science, Office of High-Energy Physics, under Contract No. DE–AC02–05CH11231, and by the National Energy Research Scientific Computing Center, a DOE Office of Science User Facility under the same contract. Additional support for DESI was provided by the U.S. National Science Foundation (NSF), Division of Astronomical Sciences under Contract No. AST-0950945 to the NSF’s National Optical-Infrared Astronomy Research Laboratory; the Science and Technology Facilities Council of the United Kingdom; the Gordon and Betty Moore Foundation; the Heising-Simons Foundation; the French Alternative Energies and Atomic Energy Commission (CEA); the National Council of Science and Technology of Mexico (CONACYT); the Ministry of Science and Innovation of Spain (MICINN), and by the DESI Member Institutions: \url{https://www.desi.lbl.gov/collaborating-institutions}. 

The DESI Legacy Imaging Surveys consist of three individual and complementary projects: the Dark Energy Camera Legacy Survey (DECaLS), the Beijing-Arizona Sky Survey (BASS), and the Mayall z-band Legacy Survey (MzLS). DECaLS, BASS and MzLS together include data obtained, respectively, at the Blanco telescope, Cerro Tololo Inter-American Observatory, NSF’s NOIRLab; the Bok telescope, Steward Observatory, University of Arizona; and the Mayall telescope, Kitt Peak National Observatory, NOIRLab. NOIRLab is operated by the Association of Universities for Research in Astronomy (AURA) under a cooperative agreement with the National Science Foundation. Pipeline processing and analyses of the data were supported by NOIRLab and the Lawrence Berkeley National Laboratory. Legacy Surveys also uses data products from the Near-Earth Object Wide-field Infrared Survey Explorer (NEOWISE), a project of the Jet Propulsion Laboratory/California Institute of Technology, funded by the National Aeronautics and Space Administration. Legacy Surveys was supported by: the Director, Office of Science, Office of High Energy Physics of the U.S. Department of Energy; the National Energy Research Scientific Computing Center, a DOE Office of Science User Facility; the U.S. National Science Foundation, Division of Astronomical Sciences; the National Astronomical Observatories of China, the Chinese Academy of Sciences and the Chinese National Natural Science Foundation. LBNL is managed by the Regents of the University of California under contract to the U.S. Department of Energy. The complete acknowledgments can be found at \url{https://www.legacysurvey.org/}. 

Any opinions, findings, and conclusions or recommendations expressed in this material are those of the author(s) and do not necessarily reflect the views of the U. S. National Science Foundation, the U. S. Department of Energy, or any of the listed funding agencies. 

The authors are honored to be permitted to conduct scientific research on Iolkam Du'ag (Kitt Peak), a mountain with particular significance to the Tohono O'odham Nation.

\software{Numpy \citep{5725236}, Scipy \citep{4160250}, Matplotlib \citep{4160265}, Astropy \citep{2013A&A...558A..33A,2018AJ....156..123A,2022ApJ...935..167A}, emcee \citep{2013PASP..125..306F}, corner \citep{2016JOSS....1...24F}, Corrfunc \citep{2020MNRAS.491.3022S}}

\section*{Data Availability}
Data points for each figure are available in a machine-readable form: \url{https://doi.org/10.5281/zenodo.7912816}

\appendix
\begin{figure*}
	\centering
	\includegraphics[scale=0.45]{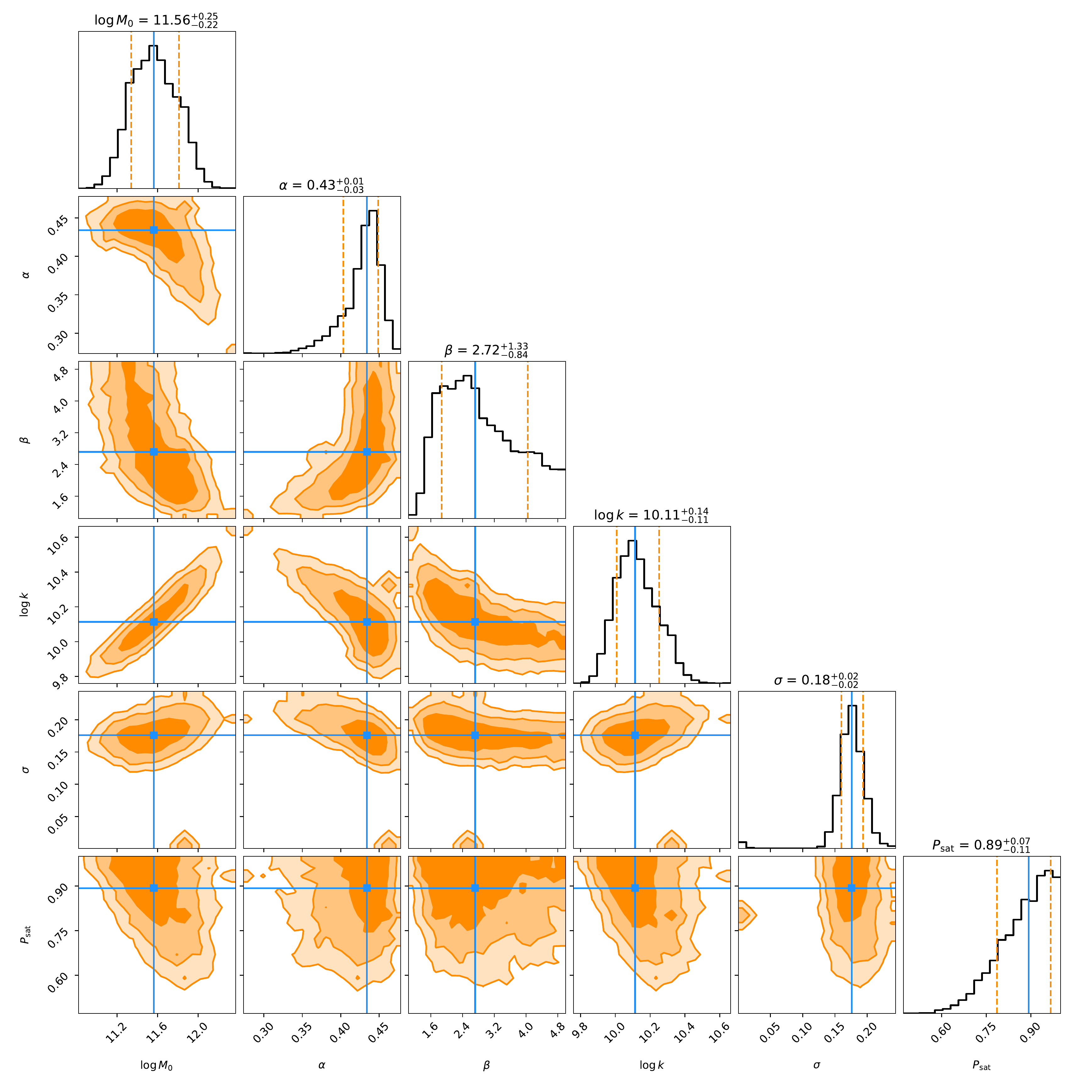}
	\caption{The posterior distributions of the parameters in the SHMR and the constant $P_{\mathrm{sat}}$ fitted with $\boldsymbol{w}^{\mathrm{obs}}_{\mathrm{p},\mathrm{LRG}}$, $\boldsymbol{w}^{\mathrm{obs}}_{\mathrm{p},\mathrm{ELG}}$ and $n^{\mathrm{obs}}_{\mathrm{LRG}}$. The contours show the joint distribution of each pair of parameters, and the three levels represent $68.3 \%$, $95.4 \%$ and $99.7 \%$ confidence intervals. The median and $1\sigma$ uncertainty derived from the marginalized distributions of each parameter are presented on the top panel of each column.
		\label{fig:corner_auto}}
\end{figure*}

\begin{figure*}
	\centering
	\includegraphics[scale=0.45]{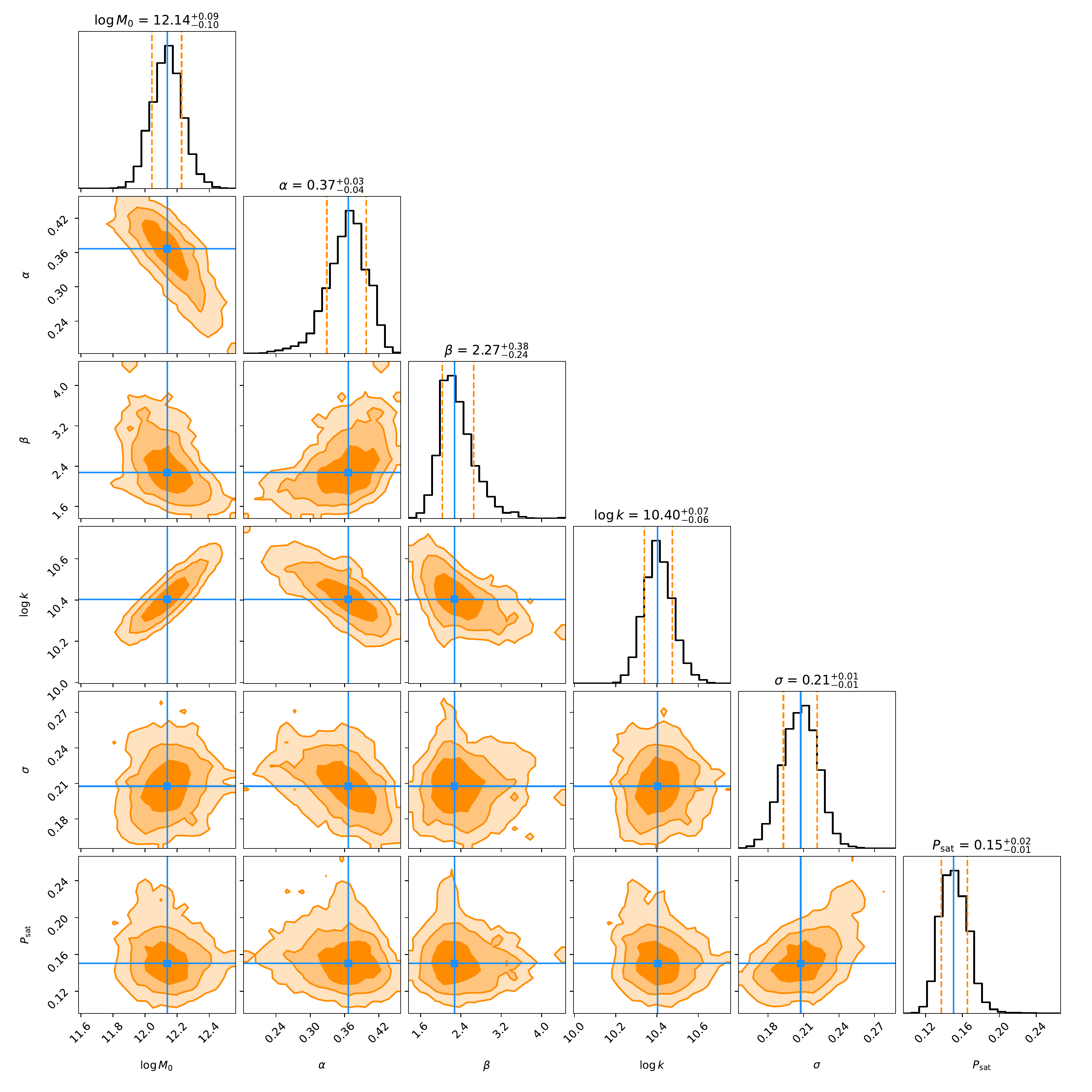}
	\caption{Similar to Figure \ref{fig:corner_auto}, but for the posterior distributions of the parameters in the SHMR and the constant $P_{\mathrm{sat}}$ fitted with $\boldsymbol{w}^{\mathrm{obs}}_{\mathrm{p},\mathrm{LRG}}$, $\boldsymbol{w}^{\mathrm{obs}}_{\mathrm{p},\mathrm{LRGxELG}}$ and $n^{\mathrm{obs}}_{\mathrm{LRG}}$.
		\label{fig:corner_cross}}
\end{figure*}

\begin{figure*}
	\centering
	\includegraphics[scale=0.4]{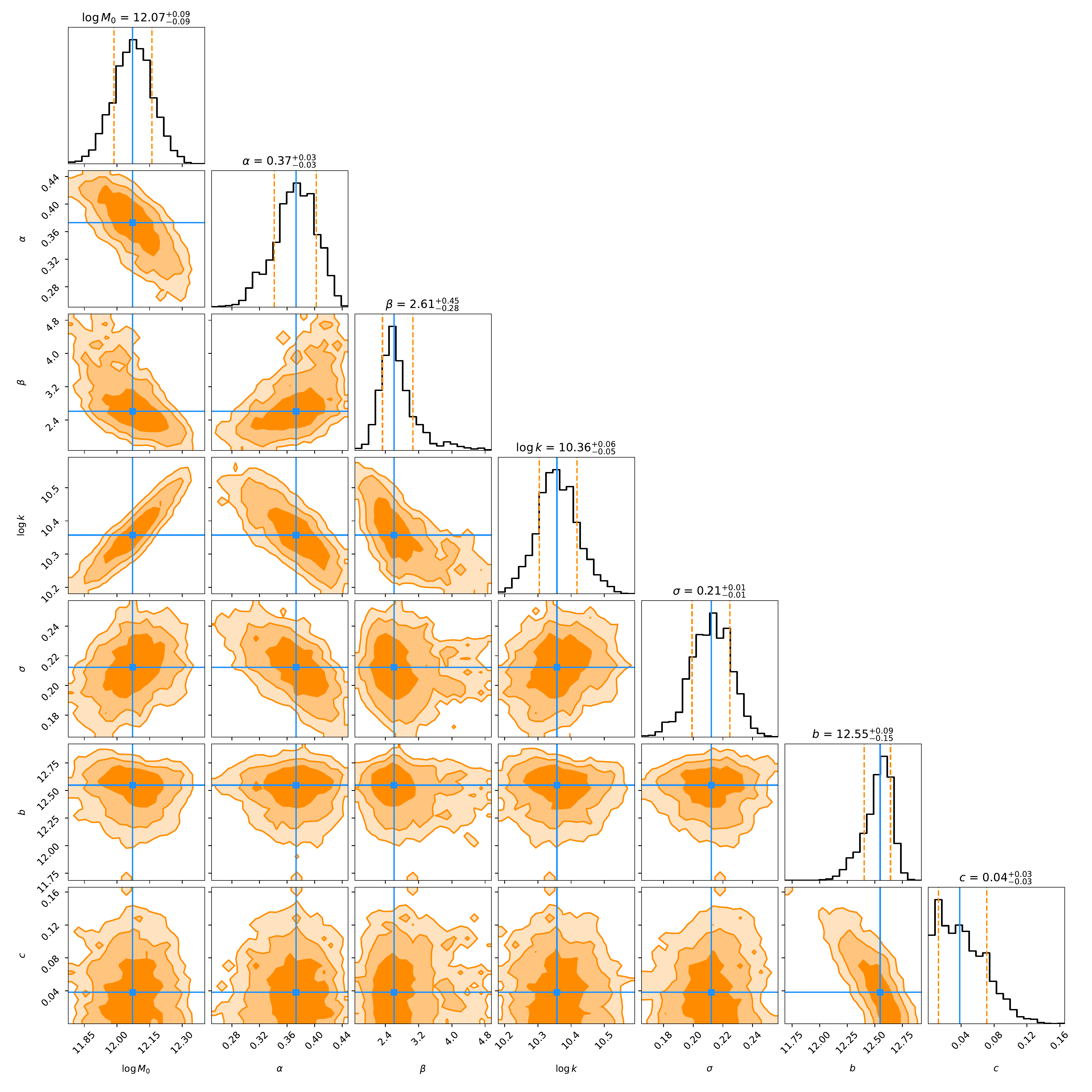}
	\caption{Similar to Figure \ref{fig:corner_auto} and \ref{fig:corner_cross}, but for the posterior distributions of the parameters in the SHMR and the halo mass-dependent $P_{\mathrm{sat}}$ model fitted with all the cross (auto) correlations $\boldsymbol{w}^{\mathrm{obs}}_{\mathrm{p},\mathrm{LRG}}$, $\boldsymbol{w}^{\mathrm{obs}}_{\mathrm{p},\mathrm{ELG}}$ and $\boldsymbol{w}^{\mathrm{obs}}_{\mathrm{p},\mathrm{LRGxELG}}$, and the LRG number densities $n^{\mathrm{obs}}_{\mathrm{LRG}}$.
		\label{fig:corner}}
\end{figure*}
\section{Posterior distributions of the model parameters}\label{sec:posterior}
We present the posterior PDFs of the parameters of the SHMR and $P_{\mathrm{sat}}$ models in this Section. Figure \ref{fig:corner_auto} and \ref{fig:corner_cross} denote the results with the constant $P_{\mathrm{sat}}$ model. The parameter constraints of the halo mass-dependent $P_{\mathrm{sat}}$ are displayed in Figure \ref{fig:corner}.

\bibliography{sv3_elg}{}
\bibliographystyle{aasjournal}

\end{document}